\documentclass[11pt,prd,onecolumn,amsmath,amssymb,aps,floats,floatfix, nofootinbib]{revtex4-2}

\usepackage[colorlinks=true,urlcolor=blue,anchorcolor=blue,citecolor=blue,filecolor=blue,linkcolor=blue,menucolor=blue,linktocpage=true]{hyperref} 

\usepackage[inline]{enumitem}
\usepackage[multidot]{grffile}  
\usepackage{dcolumn}
\usepackage{bm}
\usepackage{amsmath}
\usepackage{amsfonts}
\usepackage{amssymb}
\usepackage{color}
\usepackage[table]{xcolor}
\usepackage{float}
\usepackage{latexsym}
\usepackage{slashed} 
\usepackage{pstricks}
\usepackage{indentfirst}
\usepackage{mathrsfs}
\usepackage{multirow}
\usepackage{epsfig,psfrag}
\usepackage{graphicx}
\usepackage{enumitem}
\usepackage{geometry,amssymb,yfonts}
\usepackage{yhmath}
\usepackage{anysize}
\usepackage{subfigure}
\usepackage{mathtools}
\usepackage{setspace} 
\usepackage[utf8]{inputenc} 
\usepackage[scientific-notation=true]{siunitx} 
\usepackage{feynmp-auto}
\usepackage{verbatim}

\usepackage[normalem]{ulem}
\usepackage{longtable}

\graphicspath{{fig/}}

\allowdisplaybreaks 

\newcommand{\Uone}{\mathrm{U}(1)}

\newcommand{\UoneD}{\mathrm{U}(1)_\mathrm{D}}
\newcommand{\UoneY}{\mathrm{U}(1)_\mathrm{Y}}

\newcommand{\SUtwoD}{\mathrm{SU}(2)_\mathrm{D}}

\begin{document}

\title{Constraining the Secluded and Catalyzed Annihilation Dark Matter with Fermi-LAT and Planck Data}

\author{Yu-Hang Su$^{a}$}
\author{Chengfeng Cai$^{b}$}\email[Corresponding author. ]{caichf3@mail.sysu.edu.cn}
\author{Hong-Hao Zhang$^{a}$}\email[Corresponding author. ]{zhh98@mail.sysu.edu.cn}

\affiliation{$^a$School of Physics, Sun Yat-sen University, Guangzhou 510275, China}
\affiliation{$^b$School of Science, Sun Yat-Sen University, Shenzhen 518107, China}

\begin{abstract}
We propose a dark matter (DM) model with a complex scalar charged under a hidden gauge symmetry, denoted as $\UoneD$. The scalar field is the DM candidate while the $\UoneD$ gauge field $A'$ plays the role of a mediator, which connects the dark sector to the standard model (SM) sector via a tiny kinetic mixing. We find that both the secluded and catalyzed annihilation scenarios can be realized in this model. The phenomenology of DM, including relic density, indirect detection (Fermi-LAT), and CMB (Planck) constraints, is discussed. We also extend our discussion to DM with other spins, including Dirac fermion and vector boson. Our analysis is carried out in two models, denoted as $\UoneD \times \UoneY$ and $\UoneD \times \Uone_{L_\mu-L_\tau}$, with the former corresponding to $A'$ kinetically mixing with the $\UoneY$ gauge field $B$ and the latter corresponding to $A'$ mixing with the $\Uone_{L_\mu-L_\tau}$ gauge field $Z'$. In previous studies of catalyzed annihilation scenarios, we find that the indirect detection limits were overly restrictive because they only considered the simplified $2\mathrm{DM} \to 2\mathrm{SM}$ annihilation channel. In contrast, by performing a complete calculation of the gamma-ray and CMB constraints from the process $2\mathrm{DM} \to 2A' \to 4\mathrm{SM}$ in the models we consider, we observe weaker constraints in both the $\UoneD \times \UoneY$ and $\UoneD \times \Uone_{L_\mu-L_\tau}$ models, with the $\UoneD \times \Uone_{L_\mu-L_\tau}$ model being subject to the weakest constraints overall since it involves fewer hadronic decay processes.

\end{abstract}

\maketitle
\tableofcontents

\section{Introduction}

Dark matter (DM) is believed to constitute about $27\%$ of total energy density of our universe, supported by numerous astrophysical and cosmological evidence despite its unknown nature~\cite{Jungman:1995df, Bertone:2004pz, Feng:2010gw, Bauer:2017qwy}.
It is well known that Weakly Interacting Massive Particles (WIMPs), with masses from GeV to TeV scale and couplings similar to the weak interaction, are promising candidates for DM, since they can naturally reproduce the correct relic abundance via thermal freeze-out in the early universe~\cite{Lee:1977ua}. Various direct detection experiments have been developed to search for WIMPs, such as XENONnT~\cite{XENON:2023cxc}, PandaX-4T~\cite{PandaX:2024qfu} and LUX-ZEPLIN (LZ)~\cite{LZ:2024zvo}. 
Since no conclusive signals have been confirmed so far, direct searches place increasingly stringent constraints on WIMPs~\cite{Arcadi:2017kky, Roszkowski:2017nbc}. Therefore, many models beyond typical WIMPs have been proposed, such as Strongly Interaction Massive Particles (SIMPs)~\cite{Hochberg:2014dra, Hochberg:2014kqa, Smirnov:2020zwf}, Forbidden and Not-Forbidden DM~\cite{DAgnolo:2015ujb, Cline:2017tka, Fitzpatrick:2020vba}, Secluded and Catalyzed DM~\cite{Pospelov:2007mp, Pospelov:2008zw, Pospelov:2008jd, Xing:2021pkb, Cai:2021wmu}, and so on.

For the secluded DM model, the DM particles mainly annihilate into some lighter non-SM mediators, denoted as $A'$, which then decay into SM particles after DM freeze-out. In this case, the $2 \to 2$ DM annihilation processes are accomplished within the dark sector, so the annihilation strength is no longer correlated with the scattering strength between DM and nucleons. In particular, if the mediator is long-lived and its coupling to DM is sufficiently strong, the $3A' \to 2\mathrm{DM}$ process would play an important role when it is allowed kinematically. In this case, a new process, called catalyzed annihilation~\cite{Xing:2021pkb,Cai:2021wmu}, occurs: three $2 \mathrm{DM} \to 2A'$ processes accompanied with two $3A' \to 2 \mathrm{DM}$ processes, effectively leading to a $6\mathrm{DM} \to 4\mathrm{DM}$ process. In this scenario, the yield of DM decreases in a manner of $x^{-3/2}$ rather than $e^{-x}$. This leads to a much lower freeze-out temperature of DM comparing to the usual WIMP scenario. On the other hand, if the mass relation is inverse, $m_{A'} > m_{\chi}$, it becomes the Forbidden or Not-Forbidden DM scenario. Particularly, Refs.~\cite{Cline:2017tka, Fitzpatrick:2020vba} also discuss different freeze-out pictures when $\Gamma_{A'}$ ranges large to small. Note that the (Not-)Forbidden DM models mainly focus on sub-GeV DM, while the secluded and catalyzed DM models study the scale from GeV to TeV, which is also the mass range for the typical WIMPs DM models.

In this work, we propose a complex scalar DM model involving a hidden gauge symmetry, $\UoneD$. Complex scalar DM models have been discussed in many previous studies~\cite{Berlin:2014tja, Wu:2016mbe, Chen:2019pnt, Lao:2020inc, Su:2024oml}. We consider a complex scalar $\Phi$ charged under $\UoneD$ as the DM candidate while the gauge boson $A'$ of $\UoneD$ mediates interactions between the dark and SM sectors via the kinetic mixing. Both secluded and catalyzed scenarios can be realized in this model. We consider the DM phenomenology including relic density, gamma-ray bounds from Fermi-LAT, and CMB bounds from Planck experiment. The Fermi-LAT constraints are derived from a combined analysis of 14.3 years of observations on 42 dwarf galaxies, while the CMB constraints are based on the Planck 2018 results. For comprehensiveness, we also extend our discussion of DM phenomenology to the fermionic and vector DM models. In most of the previous studies, indirect detection constraints were usually obtained by considering a single $2\mathrm{DM}\to 2\mathrm{SM}$ annihilation channel (e.g., $W^+ W^-$ or $b\bar{b}$). However, this treatment is oversimplified in the secluded or catalyzed annihilation scenarios, leading to an overly stringent limit on the parameter space. Therefore, a goal of this work is to perform a more careful determination of the latest Fermi-LAT and CMB constraints, based on a full analysis of $2\mathrm{DM} \to 2A'\to 4\mathrm{SM}$ channels. These calculations are carried out under two different models, denoted as $\UoneD \times \UoneY$ and $\UoneD \times \Uone_{L_\mu-L_\tau}$. The main difference between these models is the portal connecting the dark sector to the SM sector. We present the phenomenological constraints on these two models and apply them to the secluded and catalyzed annihilation scenarios for DM with different spins.

It is worth noting that some previous studies have also discussed the indirect detection constraints for the secluded DM based on the $2\mathrm{DM}\to2A'\to4\mathrm{SM}$ channels~\cite{Profumo:2017obk, Siqueira:2019wdg, Siqueira:2021lqj, NFortes:2022dkj}. However, they mainly focus on single annihilation channels such as $4e$, $4\mu$, $4\tau$, or others. In contrast, the two models we consider, involve multiple decay channels of the mediator $A'$, resulting in different gamma-ray spectra of final state particles.

This paper is organized as follows. In Sect.~\ref{model}, we introduce the secluded and catalyzed DM with different spins, including complex scalar, Dirac fermion and vector boson. In Sect.~\ref{phenomenology}, we introduce the $\UoneD \times \UoneY$ and $\UoneD \times \Uone_{L_\mu-L_\tau}$ models and how we derive the constraints from Fermi-LAT and Planck. {In Sect.~\ref{results}, we present the limits and allowed parameter space for each DM spin. In the end, we draw our conclusions in Sect.~\ref{conclusions}.}

\section{Dark Matter Models}
\label{model}
\begin{figure}[!h]
\centering
\includegraphics[width=0.44\textwidth]{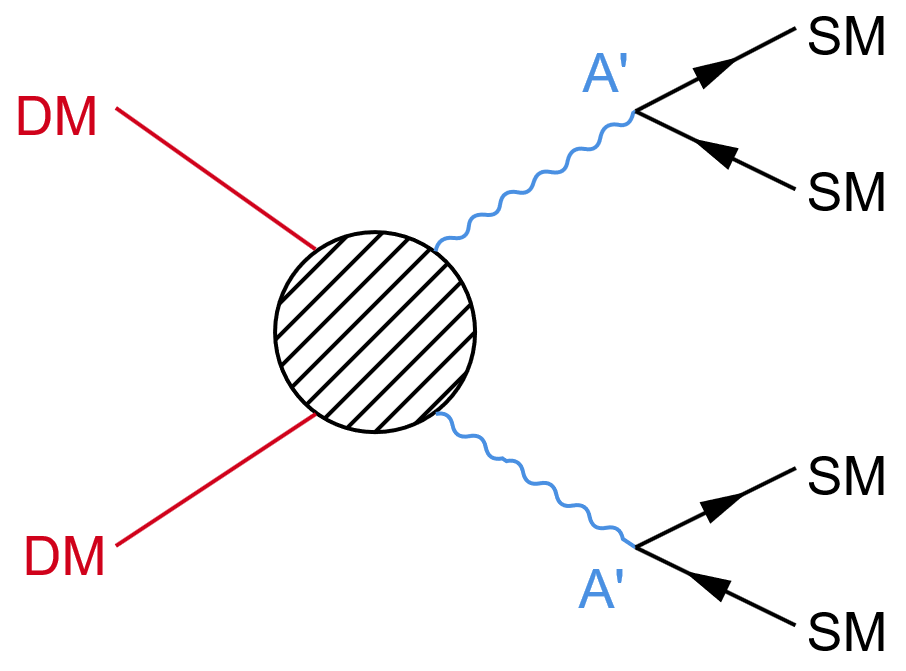}
\caption{ Feynman diagram for DM annihilating into mediator $A'$, which then decays into SM particles.}
\label{fig1}
\end{figure}

\begin{figure}[!h]
\centering
\includegraphics[width=0.9\textwidth]{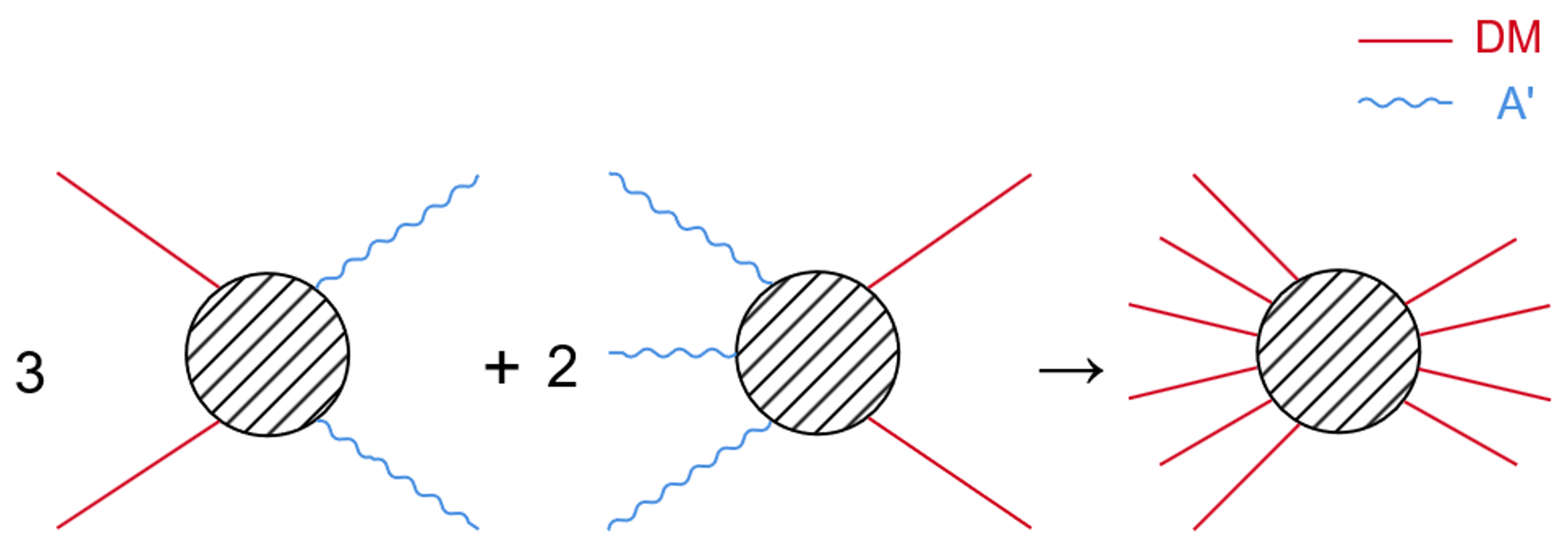}
\caption{ Feynman diagrams for catalyzed annihilation of DM with catalyst $A'$. The reduction of DM is realized through three $2\mathrm{DM} \to 2A'$ plus two $3A' \to 2\mathrm{DM}$, eventually resulting in a $6\mathrm{DM} \to 4\mathrm{DM}$ process.}
\label{fig2}
\end{figure}

In this section, we investigate the secluded and catalyzed models for DM with different spins: 0 (complex scalar), 1/2 (Dirac fermion), and 1 (vector boson). We focus on the cases that DM interacts with SM particles via vector portal. The mediator is a gauge boson denoted by $A'$, and dark matters deplete through the process $2\mathrm{DM} \to 2A' \to 4\mathrm{SM}$ (see Fig.~\ref{fig1}), and thus the mass of $A'$ should be smaller than the DM mass, i.e., $m_{A'} \lesssim m_{\chi}$. If $A'$ decays promptly when they are created, the situation would be very similar to the traditional WIMPs paradigm, except that the relic density of DM is determined by the DM-$A'$ coupling which is free from the constraint of direct detection. This scenario will be called secluded DM model in this work. On the other hand, if the decay width of $A'$, denoted as $\Gamma_{A'}$, is extremely small, it will lead to the catalyzed annihilation DM scenario, which will be called catalyzed model for short.~\footnote{Strictly speaking, the catalyzed model can be considered a type of secluded model since its DM particle is also hidden in the dark sector. However, to make a distinction in this paper, we refer to the case that only considers the $2 \to 2$ process as secluded, while the case that simultaneously considers the $3 \to 2$ process is referred to as catalyzed model.}
Fig.~\ref{fig2} illustrates how the catalyzed annihilation mechanism works to achieve the depletion of DM.

Studies of model building and DM relic density for the fermionic and vector DM have been provided in~\cite{Xing:2021pkb} and~\cite{Cai:2021wmu}, respectively. The constraints from indirect detection are also discussed in these papers, however, their analysis is overly simplified. Therefore, in this work, we will offer a detailed discussion on the setup of complex scalar DM model,  and provide a more accurate and comprehensive treatment of indirect detection constraints.

\subsection{Complex Scalar DM}
\label{model1}
The simplest model for complex scalar DM is the $\UoneD$ dark photon model, where DM $\Phi$ is charged under $\UoneD$ and $A'$ is the $\UoneD$ gauge boson.
The Lagrangian for the dark sector is given by,
\begin{eqnarray}\label{LD0}
\mathcal{L}_{D}^{(0)} = -\frac{1}{4}F'^{\mu\nu}F'_{\mu\nu} + \frac{1}{2}m_{A'}^2 A'^\mu A'_\mu + (D^\mu \Phi)^\dagger D^\mu \Phi - m^2_\Phi |\Phi|^2 - \lambda_\Phi |\Phi|^4 - \lambda_{\Phi H}|\Phi|^2 |H|^2,
\end{eqnarray}
where the covariant derivative is defined as $D_\mu = \partial_\mu - ig_D A'_\mu$. The superscript $(0)$ in $\mathcal{L}_D^{(0)}$ denotes the spin of DM. The mass of $A'$ can be generated through the Brout-Englert-Higgs mechanism or the Stueckelberg mechanism. Since we only focus on the vector-portal process, we assume $\lambda_{H\Phi}$ to be negligible, and thus the Higgs-portal annihilation or scattering processes are suppressed. Otherwise, the model would no longer be secluded but instead become the normal WIMP paradigm.

When $m_{A'} \lesssim m_{\chi}$, the dominant annihilation channels are: $2\Phi \to 2A'$ and $3A' \to 2\Phi$. The Boltzmann equations of $\Phi$ and $A'$ are given by,~\footnote{Pay attention to the different conventions for $\langle\sigma_2 v\rangle$ and $\langle\sigma_3 v^2\rangle$ used in~\cite{Cline:2017tka, Xing:2021pkb, Cai:2021wmu}. For example, Ref.~\cite{Cline:2017tka} includes the symmetry factor for initial identical particles in the Boltzmann equation, while Ref.~\cite{Cai:2021wmu} absorbs it into the cross sections. In this work, we adopt the convention used in~\cite{Cai:2021wmu}.}
\begin{eqnarray}\label{Boltzmann}
\frac{d n_\Phi}{dt} + 3H n_\Phi &=& -\frac{1}{2}\langle\sigma_2 v\rangle \left(n_\Phi^2 - \bar{n}_\Phi^2 \frac{n_{A'}^2}{ \bar{n}_{A'}^2}\right) + 2\langle\sigma_3 v^2\rangle\left(n_{A'}^3 - \bar{n}_{A'}^3 \frac{n_\Phi^2}{\bar{n}_\Phi^2} \right),\\
\frac{d n_{A'}}{dt} + 3H n_{A'} &=& \frac{1}{2} \langle\sigma_2 v\rangle\left(n_\Phi^2-\bar{n}_\Phi^2\frac{n_{A'}^2}{\bar{n}_{A'}^2} \right) -3\langle\sigma_3 v^2\rangle\left(n_{A'}^3 -\bar{n}_{A'}^3\frac{n_\Phi^2}{\bar{n}_\Phi^2} \right) - \Gamma_{A'}(n_{A'}-\bar{n}_{A'}),
\end{eqnarray}
where $\langle\sigma_2 v\rangle$ and $\langle\sigma_3 v^2\rangle$ are the thermally
averaged cross sections of $2\Phi \to 2A'$ and $3A' \to 2\Phi$, respectively. $n_\Phi$ represents the total number density of DM $\Phi$ and anti-DM $\Phi^\dagger$, while $\bar{n}_{\Phi, A'}$ denote the equilibrium densities.  

\begin{figure}[!h]
\centering
\includegraphics[width=0.8\textwidth]{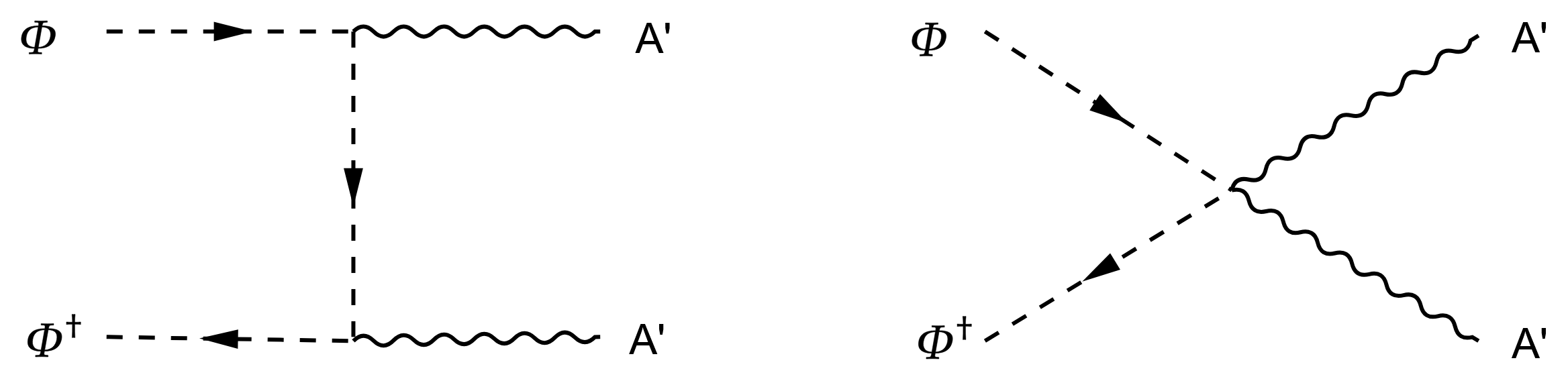}
\caption{ Feynman diagrams of $2\Phi \to 2A'$ process. Note that the first plot includes t- and u-channels, which can be obtained by exchanging the external lines of $A'$.}
\label{fig3}
\end{figure}

\begin{figure}[!h]
\centering
\includegraphics[width=1\textwidth]{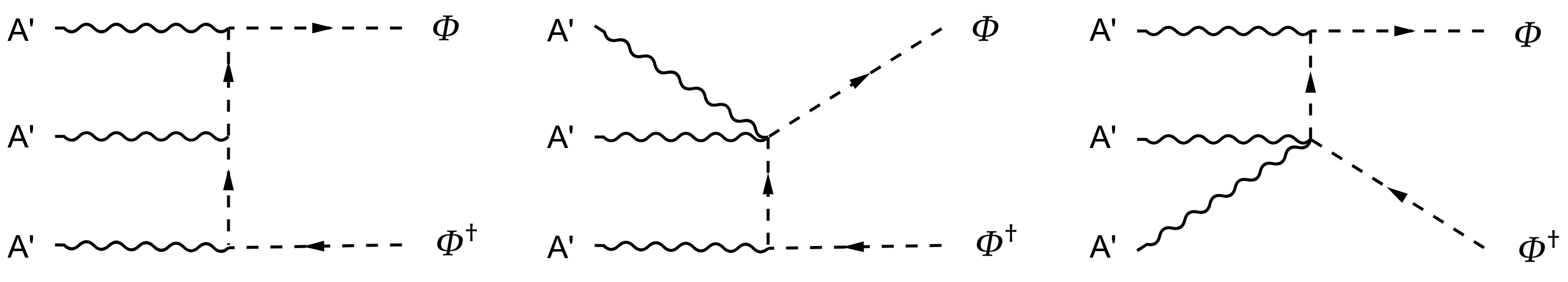}
\caption{ Feynman diagrams of $3A' \to 2\Phi$ process. Similarly, by exchanging the external lines of $A'$, the first plot has 6 independent diagrams, while the latter two each have 3 independent diagrams.}
\label{fig4}
\end{figure}

The Feynman diagrams of $2\Phi \to 2A'$ and $3A' \to 2\Phi$ are shown in Fig.~\ref{fig3} and Fig.~\ref{fig4}, respectively. In the kinematic threshold limit, which is equivalent to adopting a non-relativistic approximation keeping only the leading-order terms, their thermally averaged cross sections can be obtained using the formulas (E4) and (E5) in~\cite{Cline:2017tka}, which yield the following results:
\begin{eqnarray}\label{sv1}
\langle\sigma_2 v\rangle &=& \frac{g_D^4(8r^4-8r^2+3)}{16\pi m_\Phi^2(1-2r^2)^2} \sqrt{1-\frac{1}{r^2}}, \\
\langle\sigma_3 v^2\rangle &=& \frac{g_D^6 r^5(-64r^6 + 368r^4 - 716r^2 + 477)}{1536\pi m_\Phi^5} \sqrt{9-4r^2},
\end{eqnarray} 
where $r \equiv m_\Phi/m_{A'}$.

$A'$ can mix with other gauge bosons via kinetic term, making it unstable and causing it to eventually decay into SM particles. The simplest realization is to let $A'$ mix with the $\UoneY$ gauge field $B$:
\begin{eqnarray}\label{mixing}
\mathcal{L}_{\text{kin}} \supset -\frac{s_\epsilon}{2} F'_{\mu\nu} B^{\mu\nu},
\end{eqnarray}
where $F'^{\mu \nu}$ is the field strength tensor of $A'$. This realization, which we denote as the $\UoneD \times \UoneY$ model, can also induce scattering between DM and nuclei. The $\Phi$-proton cross section can be estimated using effective field theory~\cite{Zheng:2010js, Yu:2011by}:
\begin{eqnarray}
\sigma_{V,\Phi p} \simeq \frac{m_p^2}{\pi}\left(\frac{e c_W g_D s_\epsilon}{m_{A'}^2}\right)^2.
\end{eqnarray}
Comparing it to the latest direct detection bound from the LZ experiment~\cite{LZ:2024zvo}, $\sigma_{\mathrm{SI}} \sim 3 \times 10^{-47}~\mathrm{cm^2}$ for $1~\mathrm{TeV}$ DM, we find that the kinetic mixing angle is constrained to be:
\begin{eqnarray}
s_\epsilon \lesssim 2 \times 10^{-3} \left(\frac{1}{g_D}\right)\left(\frac{m_\Phi}{1~\mathrm{TeV}}\right)^2,
\end{eqnarray}
assuming $m_{A'} \approx m_\Phi$. In the following discussion of several benchmarks, the parameters are chosen to satisfy this bound. In addition, since $s_\epsilon \ll 1$, we can safely treat $A'$ as its mass eigenstate. The decay width of $A'$ induced by the kinetic mixing can be simply estimated as~\cite{Gabrielli:2015hua}:
\begin{eqnarray}\label{GammaAp}
\Gamma_{A'} \simeq \frac{27 \alpha s_\epsilon^2 m_{A'}}{16c_W^2} \simeq 2 \times 10^{-2} s_\epsilon^2 m_{A'}.
\end{eqnarray}

\begin{figure}[h!]
\centering
\subfigure[Catalyzed process\label{fig5-1}]
{\includegraphics[width=0.48\textwidth]{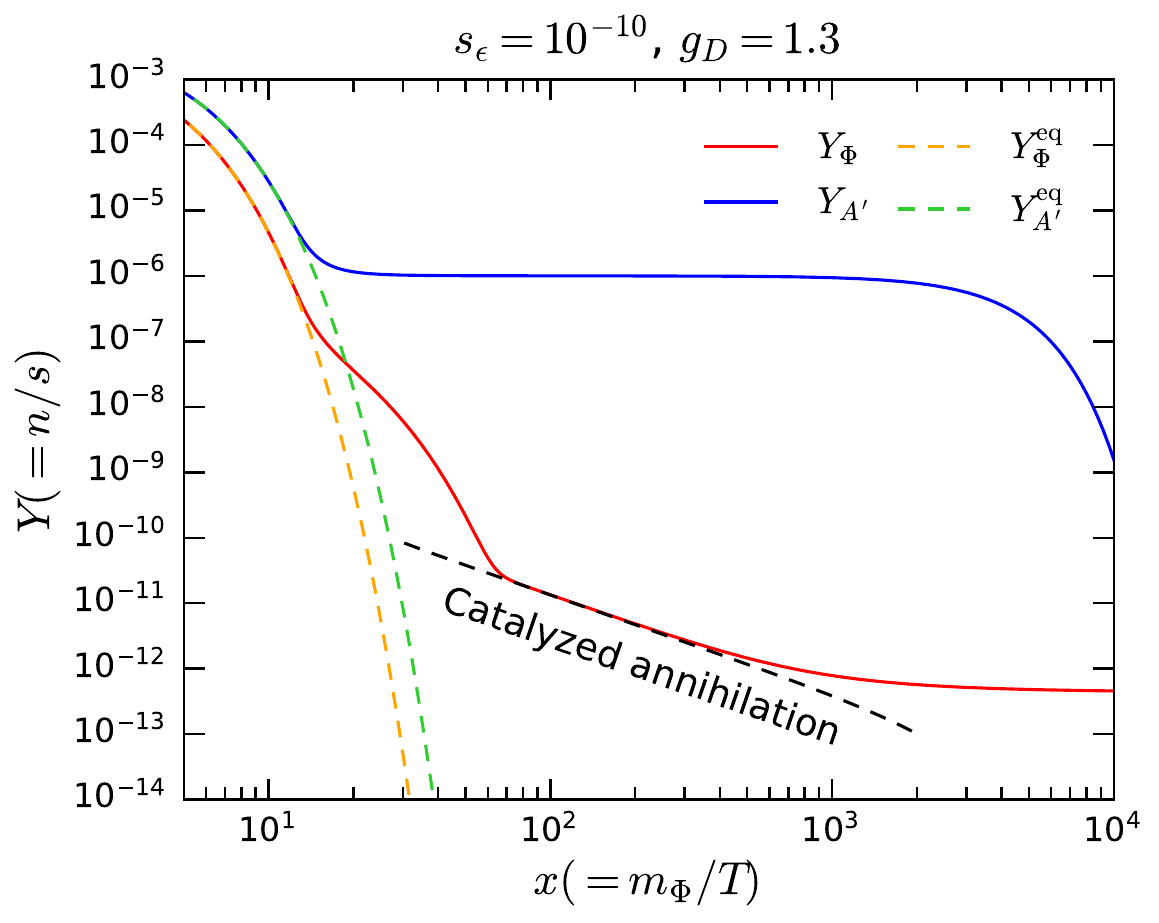}}
\hspace{.01\textwidth}
\subfigure[Semi-Catalyzed process\label{fig5-2}]
{\includegraphics[width=0.48\textwidth]{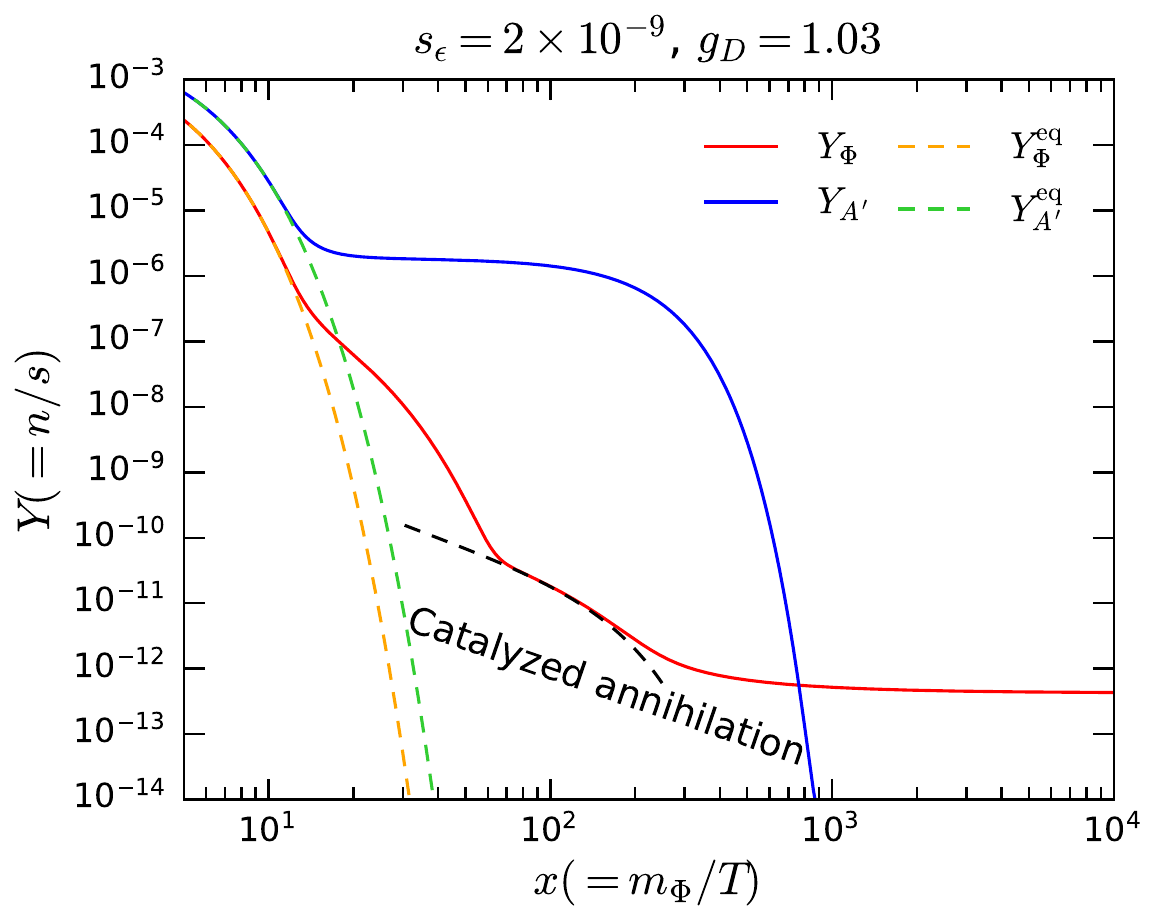}}
\hspace{.01\textwidth}
\subfigure[Secluded process\label{fig5-3}]
{\includegraphics[width=0.48\textwidth]{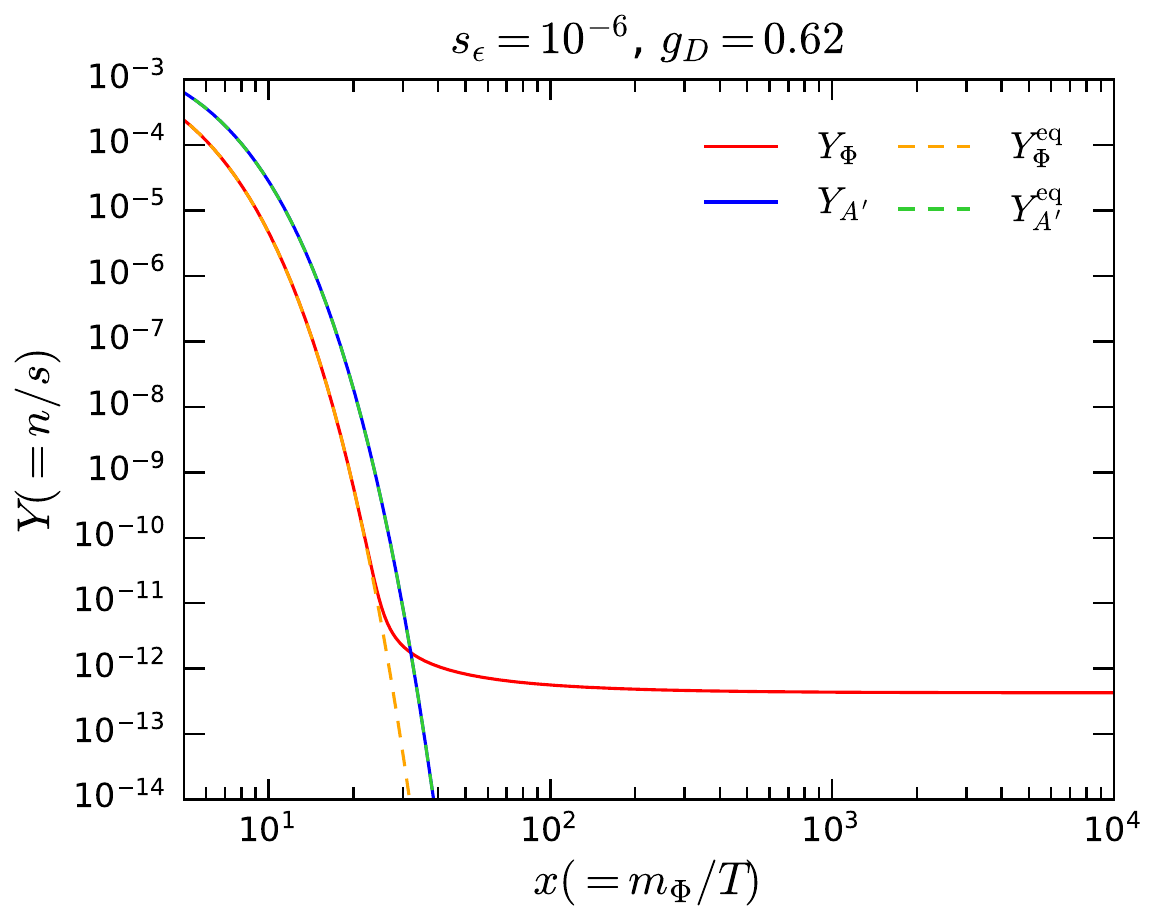}}
\caption{The evolution of the yields of DM $\Phi$ (solid red) and mediator $A'$ (solid blue) for $s_{\epsilon} = 10^{-10}$ (top left panel), $2 \times 10^{-9}$ (top right panel), and $10^{-6}$ (bottom panel). The gauge coupling $g_D$ is chosen to be $1.3$, $1.03$, and $0.62$ respectively, to obtain the correct DM relic abundance, $\Omega_{\Phi}h^2 = 0.12$. As $s_{\epsilon}$ increases, the catalyzed annihilation is terminated earlier by $A'$ decay, thus the model reduces to a secluded one eventually.}
\label{fig5}
\end{figure}

Fig.~\ref{fig5} illustrates the thermal evolution of DM $\Phi$ (solid red) and mediator $A'$ (solid blue) as the temperature decreases for benchmark models with $s_{\epsilon} = 10^{-10}$ (\ref{fig5-1}), $2 \times 10^{-9}$ (\ref{fig5-2}), and $10^{-6}$ (\ref{fig5-3}). The gauge coupling $g_D$ is adjusted to $1.3$, $1.03$, and $0.62$ respectively, in order to reproduce the observed DM relic abundance, $\Omega_{\mathrm{DM}}h^2 = 0.12$~\cite{Planck:2018vyg}. Other parameters are fixed as follows: $m_{\Phi} = 10^3~\mathrm{GeV}$ and $r = 1.2$. Fig.~\ref{fig5-1} depicts a complete catalyzed annihilation process until DM freeze-out, without termination from $A'$ decay. Fig.~\ref{fig5-2} illustrates the scenario where an increased $s_{\epsilon}$ ($\Gamma_{A'}$) leads to the sudden termination of the catalyzed annihilation processes via the decay of $A'$. This forces the DM to freeze out earlier. We refer to this as a semi-catalyzed annihilation compared to the former case. Fig.~\ref{fig5-3} corresponds to a standard $2 \to 2$ freeze-out process without catalyzed annihilation. In this case, $A'$ remains in thermal equilibrium and never freezes out, making the model reduce to a secluded one. Comparing the plots in Fig.~\ref{fig5}, we can see that $g_D$ decreases with increasing $s_{\epsilon}$ for maintaining the correct relic abundance. This can be expected since the depletion of DM in the catalyzed annihilation scenario is less efficient than the usual secluded scenario, thus a larger $g_D$ is required to enhance the annihilation rate of DM.

We note that with such a small mixing angle $s_{\epsilon}$, maintaining kinetic equilibrium (KE) between the dark sector and the SM sector becomes difficult. To address this issue, we present a method in Appendix~\ref{KE} to keep DM in KE until freeze-out, ensuring temperature equality between the dark and SM sectors.

\subsection{Dirac fermionic DM}
\label{model2}

We can also consider the DM to be a Dirac fermion, which is charged under the $\UoneD$ gauge symmetry. In this case, the Lagrangian for the dark sector becomes:
\begin{eqnarray}
\mathcal{L}_{D}^{(1/2)} &=& -\frac{1}{4}F'^{\mu\nu}F'_{\mu\nu} + \frac{1}{2}m_{A'}^2 A'^\mu A'_\mu + \bar{\chi} (i\slashed{D} - m_\chi) \chi.
\end{eqnarray}
The annihilation cross sections of $2\chi \to 2A'$ and $3A' \to 2\chi$ processes are given by~\cite{Cline:2017tka},
\begin{eqnarray}\label{sv2}
\langle\sigma_2 v\rangle &=& \frac{g_D^4(r^4-r^2)}{4\pi m_\chi^2(1-2r^2)^2} \sqrt{1-\frac{1}{r^2}}, \\
\langle\sigma_3 v^2\rangle &=& \frac{g_D^6 r^5(24r^6 - 60r^4 - 47r^2 + 153)}{2592\pi m_\chi^5} \sqrt{9-4r^2}.
\end{eqnarray}
The Boltzmann equations have the same form as Eqs.~\eqref{Boltzmann}, except that $\Phi$ is replaced by $\chi$.

\subsection{Vector Boson DM}
\label{model3}
We can also consider the DM to be a complex vector field $X_\mu$. In this case, both the DM and mediator are vector fields, and thus an elegant realization is to unify the DM and the mediator by an  $\SUtwoD$ gauge symmetry~\cite{Cai:2021wmu}. The DM $X$ and the mediator $A'$ stem from the components of the $\SUtwoD$ gauge fields, $V^a_\mu$ ($a = 1,2,3$). To be precise, the DM field is defined as $X_\mu=(V^1_\mu -iV^2_\mu)/\sqrt{2}$, while the mediator is $A'_\mu=V^3_\mu$. Their masses can be generated by the Brout-Englert-Higgs mechanism via non-zero vacuum expectation values (VEVs) of a doublet $\Phi_D$ and a real triplet $\Delta_D$ scalar fields. In this scenario, the Lagrangian for the dark sector is given by,
\begin{eqnarray}
\mathcal{L}_{D}^{(1)} &=& -\frac{1}{4} V^a_{\mu \nu} V^{a, \mu \nu} + (D^\mu \Phi_D)^\dagger D^\mu \Phi_D + \mathrm{Tr}[ (D^\mu \Delta_D)^\dagger D^\mu \Delta_D] - V(\Phi_D, \Delta_D),
\end{eqnarray}
where $V(\Phi_D, \Delta_D)$ is the potential terms of the new scalar fields. The potential is assumed to have a form such that both $\Phi_D$ and $\Delta_D$ acquire non-zero VEVs. In this situation, the DM is heavier than the mediator automatically. The cross sections of $2X \to 2A'$ and $3A' \to 2X$ processes are given by~\cite{Cai:2021wmu},
\begin{eqnarray}\label{sv3}
\langle\sigma_2 v\rangle &=& \frac{g_D^4(152r^4 - 136r^2 + 128 - 18r^{-2} + 3r^{-4})}{144\pi m_X^2(1-2r^2)^2} \sqrt{1-r^{-2}}, \\
\langle\sigma_3 v^2\rangle &=& \frac{g_D^6 r^5 f(r)}{3456\pi m_X^5} \sqrt{9-4r^2},\\
f(r) &=& -48r^6 - 12r^4 + 415r^2 - \frac{2317}{4} + \frac{2585}{4}r^{-2} - \frac{1007}{4}r^{-4} + \frac{1285}{8}r^{-6} \nonumber\\
&+& \frac{675}{16}r^{-8} - \frac{243}{16}r^{-10} + \frac{729}{256}r^{-12}.
\end{eqnarray} 

In the $\SUtwoD$ model, the kinetic mixing between $A'$ and the $\UoneY$ gauge field $B_\mu$ can be realized through a dimension-5 effective operator:
\begin{eqnarray}
\mathcal{L}_{5} &=& -\frac{c}{\Lambda} B^{\mu \nu} \Delta^a_D V^a_{\mu \nu} \supset -\frac{s_\epsilon}{2} F'_{\mu\nu} B^{\mu\nu},
\end{eqnarray}
where $c$ is a Wilson coefficient, and $\Lambda$ represents a UV complete scale. This term can generate the same kinetic mixing term as in the $\UoneD$ model. Consequently, the decay width and decay channels of $A'$ are identical to those in the $\UoneD$ model.

The Boltzmann equations for DM and mediator particles have the same form as Eqs.~\eqref{Boltzmann}, except that $\Phi$ is replaced by $X$.

\section{Dark Matter Phenomenology}
\label{phenomenology}

In this section, we investigate the astrophysical signatures arising from the annihilation of secluded or catalyzed DM. Searching for these DM directly would be difficult if the kinetic mixing parameter is extremely small, since the cross section of DM-nucleon scattering depends on the mixing directly. On the other hand, the annihilation cross section of DM is dominated by $\bar{\chi}-\chi-A'$ coupling, which can be significant. Therefore, indirect detections of DM might play a crucial role in DM searches. For instance, gamma-ray flux can be generated by the charged products originating from $A'$ decay. In this case, the production rate of $A'$, which is determined by the annihilation cross section of the $2\mathrm{DM} \to 2A'$ process, is the most relevant quantity.

In the $\UoneD \times \UoneY$ model~\footnote{For the vector DM introduced in Sect.~\ref{model3}, it should be the $\SUtwoD \times \UoneY$ model. For convenience, we use $\UoneD$ to represent the three types of DM introduced in Sect.~\ref{model} throughout this work. In Sect.~\ref{results}, we use current indirect detection bounds to constrain all these three types of DM.}
$A'$ can decay into various final states, including quark-antiquark pairs $q\bar{q}$ ($q = u, d, s, c, b, t$), charged leptons $l\bar{l}$ ($l = e, \mu, \tau$), neutrinos $\nu\bar{\nu}$ ($\nu = \nu_e, \nu_\mu, \nu_\tau $), $W^+ W^-$, and $Z$ or $h$ bosons. When we consider the annihilation chain processes $2\chi\to 2A' \to 4\text{SM}$, the situation becomes more complicated. Although the \texttt{PPPC4DM} code provides gamma-ray spectra for pure channels like $2\chi \to 2V \to 4e$, $4\mu$, $4\tau$~\cite{Cirelli:2010xx}, it cannot be directly applied to our model, since in addition to pure channels like $2\chi \to 4q$, $4l$, $4\nu$, there are also mixed channels such as $2\chi \to 2q 2l$, $2q2\nu$, $2l 2\nu$, $2qW^+ W^-$, $2qZh$, etc.. After determining the hadronization, these processes may lead to intricate distributions of the kinetic energies of final-state particles. Moreover, \texttt{PPPC4DM} assumes that the mediator is only slightly heavier than its decay products (usually much lighter than the DM), whereas in our model, $A'$ is just slightly lighter than DM. Given these considerations, we utilize the \texttt{Pythia8}~\cite{Bierlich:2022pfr} event generator to obtain the gamma-ray spectra in our work.
Moreover, since \texttt{Pythia8} now directly supports electroweak (EW) showers to simulate the soft radiation of EW gauge bosons, it allows us to incorporate EW corrections into the calculations of gamma-ray spectra and indirect detection constraints.

Since $A'$ in the $\UoneD \times \UoneY$ model can decay into hadronic final states, it produces more gamma rays comparing to pure leptonic decays. On the other hand, if we consider a model where $A'$ dominantly decays into leptons, then the constraints from indirect detections will be released. In this work, we will consider an extension of SM with $\UoneD \times \Uone_{L_\mu-L_\tau}$ gauge symmetries, and the mediator field $A'$ kinetically mixes with the gauge boson, $Z'$, corresponding to the $\Uone_{L_\mu-L_\tau}$ gauge symmetry. The $\Uone_{L_\mu-L_\tau}$ model has various potential implications. For instance, it may provide an explanation for the electron and muon $(g-2)$ anomalies~\cite{Bodas:2021fsy, Panda:2022kbn, Borah:2021khc}. Additionally, it could lead to the production of dark photons at the MUonE experiment~\cite{GrillidiCortona:2022kbq}. Moreover, it may help to interpret the excess of electrons and positrons observed in cosmic-ray measurements~\cite{Duan:2017qwj, He:2009ra}. 

The kinetic mixing term in this extended model changes from Eq.~\eqref{mixing} to:
\begin{eqnarray}\label{mixing2}
\mathcal{L}'_{\text{kin}} \supset -\frac{s'_\epsilon}{2} F'_{\mu\nu} Z'^{\mu\nu},
\end{eqnarray}
where $Z'^{\mu \nu}$ is the field strength tensor of $Z'$. The Lagrangian for the Beyond the Standard Model (BSM) sector becomes:
\begin{eqnarray}
\mathcal{L}_{BSM}^{\prime (S)} = \mathcal{L}_{D}^{(S)} - \frac{1}{4}Z'^{\mu\nu}Z'_{\mu\nu} + \frac{1}{2}m_{Z'}^2 Z'^\mu Z'_\mu + g_x J_\nu^x Z'^\nu,
\end{eqnarray}
where the superscript $(S)$ denotes the spin of DM and $g_x$ denotes the $\Uone_{L_\mu-L_\tau}$ gauge coupling. The current $J_\nu^x$ is given by,
\begin{eqnarray}
J_\nu^x = \bar{L}_\mu \gamma_\nu L_\mu + \bar{\mu}_R \gamma_\nu \mu_R - \bar{L}_\tau \gamma_\nu L_\tau - \bar{\tau}_R \gamma_\nu \tau_R.
\end{eqnarray}

In this model, $A'$ dominantly decays into leptonic final states, including $\mu$ pairs, $\tau$ pairs, and corresponding neutrino partners via the current $J^x_\nu$. As a result, $A'$ has only four decay channels in this model, making it much simpler than the $\UoneD \times \UoneY$ model. We also use \texttt{Pythia8} to obtain its gamma-ray spectra.

It is worth noting that the model-building approach for the $\UoneD \times \Uone_{L_\mu-L_\tau}$ model can be extended to other $\Uone'$ models, such as $\Uone_{L_\mu-L_e}$, $\Uone_{L_e-L_\tau}$, and $\Uone_{B-L}$. The decay channels of $A'$ vary across these models, leading to different gamma-ray spectra. Consequently, the constraints from indirect searches are model-dependent. In some previous studies, constraints of gamma rays are often derived from $2\mathrm{DM}\to2\mathrm{SM}$ processes in a single annihilation channel (e.g., $W^+ W^-$ or $b\bar{b}$). However, this treatment is oversimplified for secluded or catalyzed annihilation DM models, since the actual processes are $2\mathrm{DM}\to4\mathrm{SM}$ processes. The approach employed in this work, which considers the full decay spectra of $A'$ via numerical simulation, can derive much more accurate results.

\subsection{Gamma rays}
\label{Fermi-LAT}

DM particles can continuously self-annihilate to produce SM particles, especially in regions with high DM density. Stable products like positrons, electrons, and gamma rays can be observed by telescopes like the Fermi Large Area Telescope (Fermi-LAT), which detects gamma-ray data from the Milky Way's dwarf spheroidal galaxies (dSphs). DSphs are regarded as perfect sources for DM searches due to their proximity, DM dominance, and low astrophysical background, which minimize contamination from non-DM sources.
Since no significant DM annihilation signal from dSphs has been observed so far, these observations have placed tight constraints on the annihilation cross sections for various DM annihilation channels~\cite{Fermi-LAT:2015att, McDaniel:2023bju, MAGIC:2016xys, Profumo:2016idl}.

We utilize the Fermi-LAT 2023 data from the P8R3SOURCEV3 event class, based on 14.3 years of observations of 42 dSphs~\cite{McDaniel:2023bju}~\footnote{These dSphs belong to the "Benchmark" samples in~\cite{McDaniel:2023bju}, including 30 dSphs with measured J-factors from the "Measured" sample, plus additional 12 dSphs that only have J-factor estimates based on the kinematic or photometric scaling relation~\cite{Pace:2018tin}. 8 "Special" dSphs are excluded for several issues that could affect DM searches.}.
The data covers an energy range from $500~\mathrm{MeV}$ to $1~\mathrm{TeV}$, with likelihoods extracted for each energy bin. The collaboration has made these likelihood functions publicly available~\footnote{https://www-glast.stanford.edu/pub\_data/1841/.}, enabling researchers to perform joint likelihood analyses across multiple dSphs for any given gamma-ray spectrum.

The expected differential gamma-ray flux from DM annihilation in a given angular direction ($\Delta \Omega$) is given by~\cite{Cirelli:2010xx},
\begin{eqnarray}\label{flux}
\frac{d\Phi}{dE}(\Delta \Omega) = \frac{1}{4\pi \eta} \frac{\langle \sigma v \rangle}{m_{\text{DM}}^2} \left(\frac{dN}{dE} \right)_\gamma \cdot J,
\end{eqnarray}
where $\langle \sigma v \rangle$ is the thermal annihilation cross section of DM, and $\left(\frac{dN}{dE} \right)_\gamma$ is the average gamma-ray energy spectrum per annihilation. The factor $\eta = 2$ ($4$) corresponds to self-conjugate (non-self-conjugate) DM, and in this work, we consider non-self-conjugate DM models. The astrophysical J-factor $J$ is given by,
\begin{eqnarray}
J = \int_{\Delta \Omega} d\Omega \int_{\text{l.o.s}} ds \, \rho_{\text{DM}}^2(s),
\end{eqnarray}
where the integrals run over $\Delta \Omega$ and the line-of-sight (l.o.s.) through the DM distribution.  We employ the DM density Navarro-Frenk-White (NFW) profile here~\cite{Navarro:1996gj}, which aligns with the choice of Fermi-LAT collaboration. The profile reads,
\begin{eqnarray}
 \rho_{\text{DM}} = \frac{\rho_0 r_s^3}{r(r_s + r)^2},
\end{eqnarray}
where $\rho_0$ is the characteristic density and $r_s$ is the scale radius.
It should be noted that adopting other profiles, such as a steeper profile~\cite{Graham:2005xx} or a more core-like profile~\cite{Burkert:1995yz, Salucci:2000ps}, would result in a shift of our limits by a constant factor.

\begin{figure}[!h]
\centering
\subfigure[\label{spec1-1}]
{\includegraphics[width=0.48\textwidth]{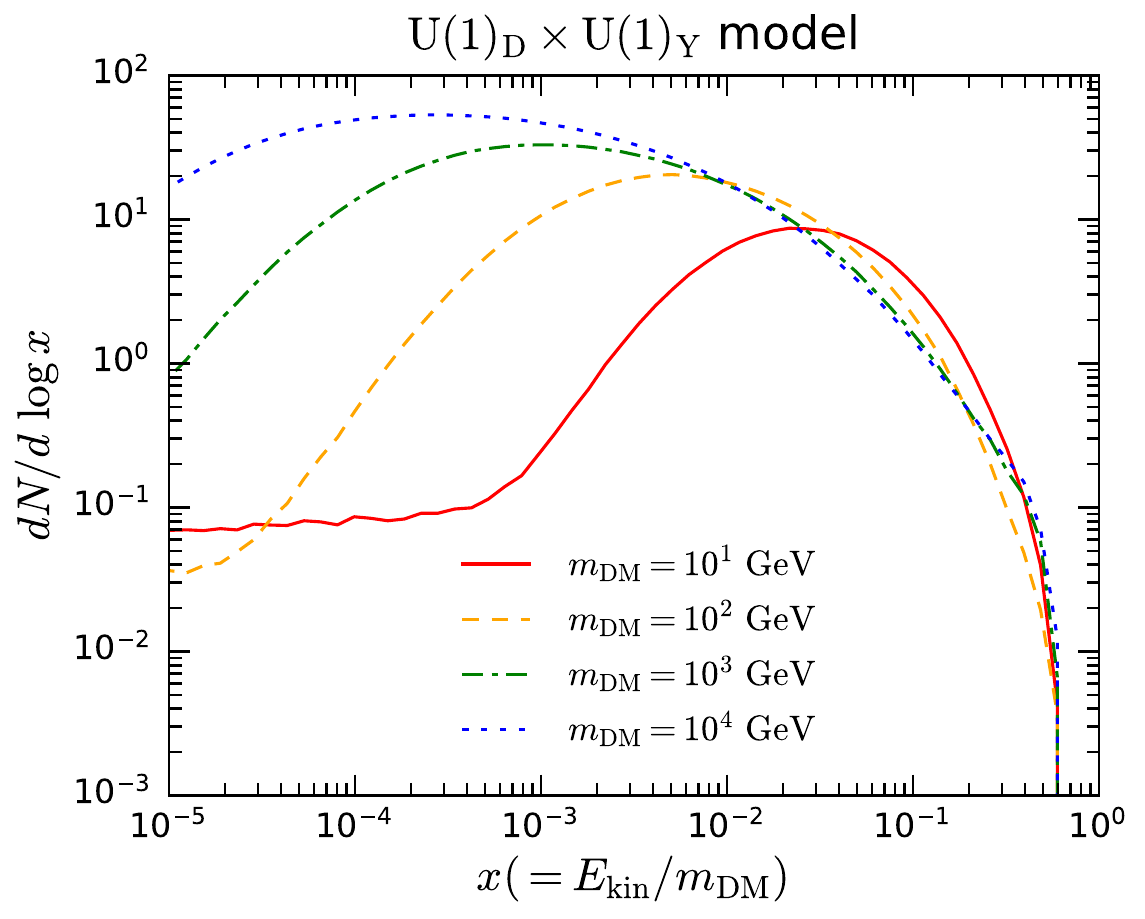}}
\hspace{.01\textwidth}
\subfigure[\label{spec1-2}]
{\includegraphics[width=0.48\textwidth]{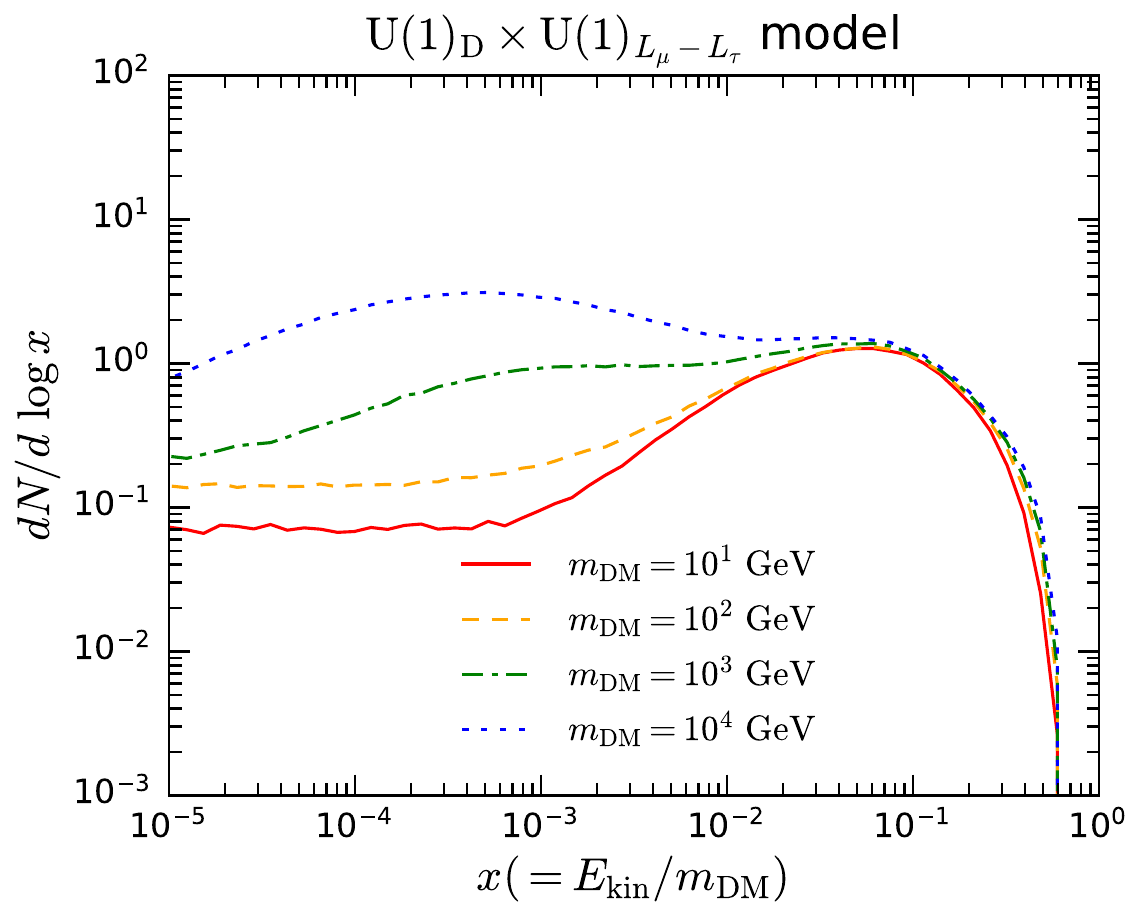}}
\caption{Gamma-ray spectra for DM annihilates into $A'A'$ for various DM masses: $10~\mathrm{GeV}$ (red), $10^2~\mathrm{GeV}$ (orange), $10^3~\mathrm{GeV}$ (green), and $10^4~\mathrm{GeV}$ (blue).  The left panel presents results for the $\UoneD \times \UoneY$ model, while the right panel shows those for the $\UoneD \times \Uone_{L_\mu-L_\tau}$ model.}
\label{spec1}
\end{figure}

In the secluded and catalyzed annihilation scenarios, the model-dependent part in Eq.~\eqref{flux} is the energy spectrum $(dN/dE)_\gamma $, which can be computed by \texttt{Pythia8}. In Fig.~\ref{spec1}, we present gamma-ray spectra for the $\UoneD \times \UoneY$ and $\UoneD \times \Uone_{L_\mu-L_\tau}$ models, where different color lines represent different DM masses. The $\UoneD \times \UoneY$ model (left panel) produces more gamma rays compared to the $\UoneD \times \Uone_{L_\mu-L_\tau}$ model (right panel) since the mediator $A'$ decaying into quarks and bosons can produce more gamma rays compared to leptonic final states~\cite{Cirelli:2010xx}.

\begin{figure}[!h]
\centering
\subfigure[\label{spec2-1}]
{\includegraphics[width=0.48\textwidth]{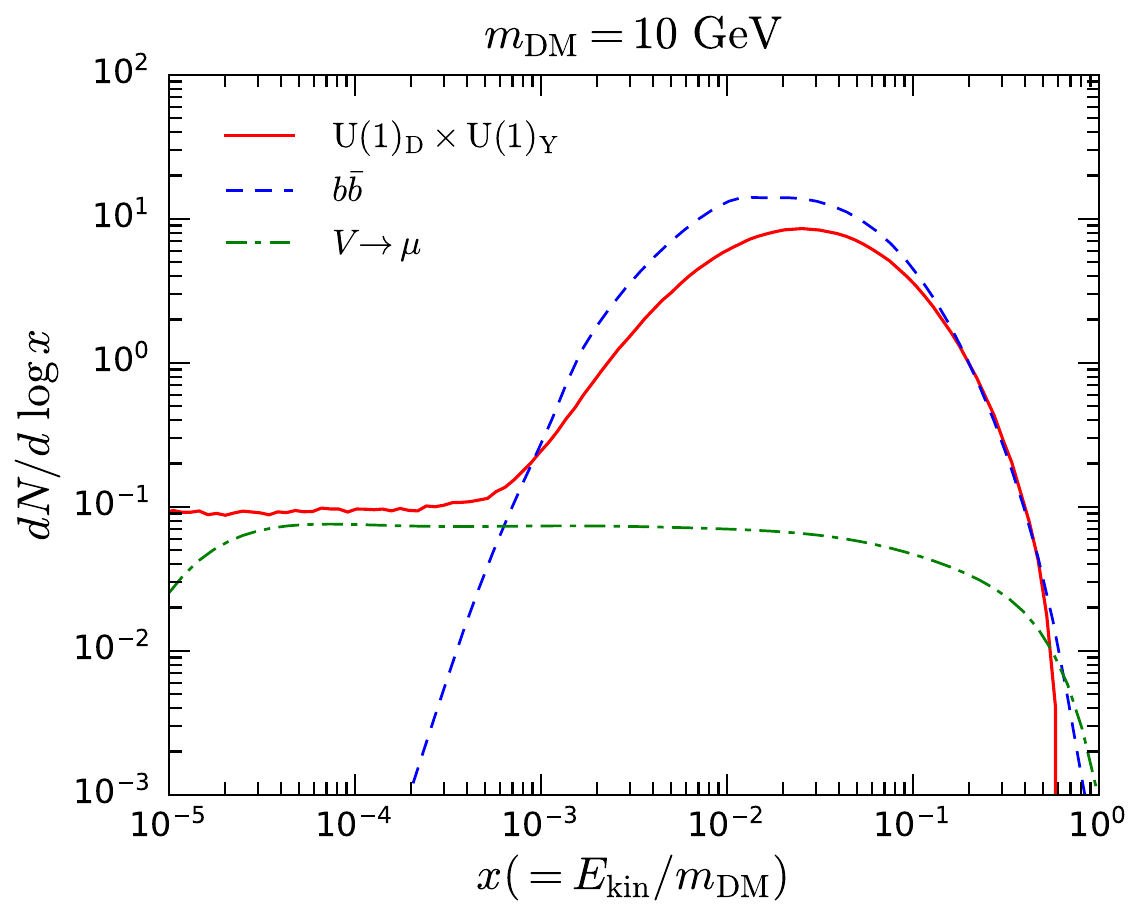}}
\hspace{.01\textwidth}
\subfigure[\label{spec2-2}]
{\includegraphics[width=0.48\textwidth]{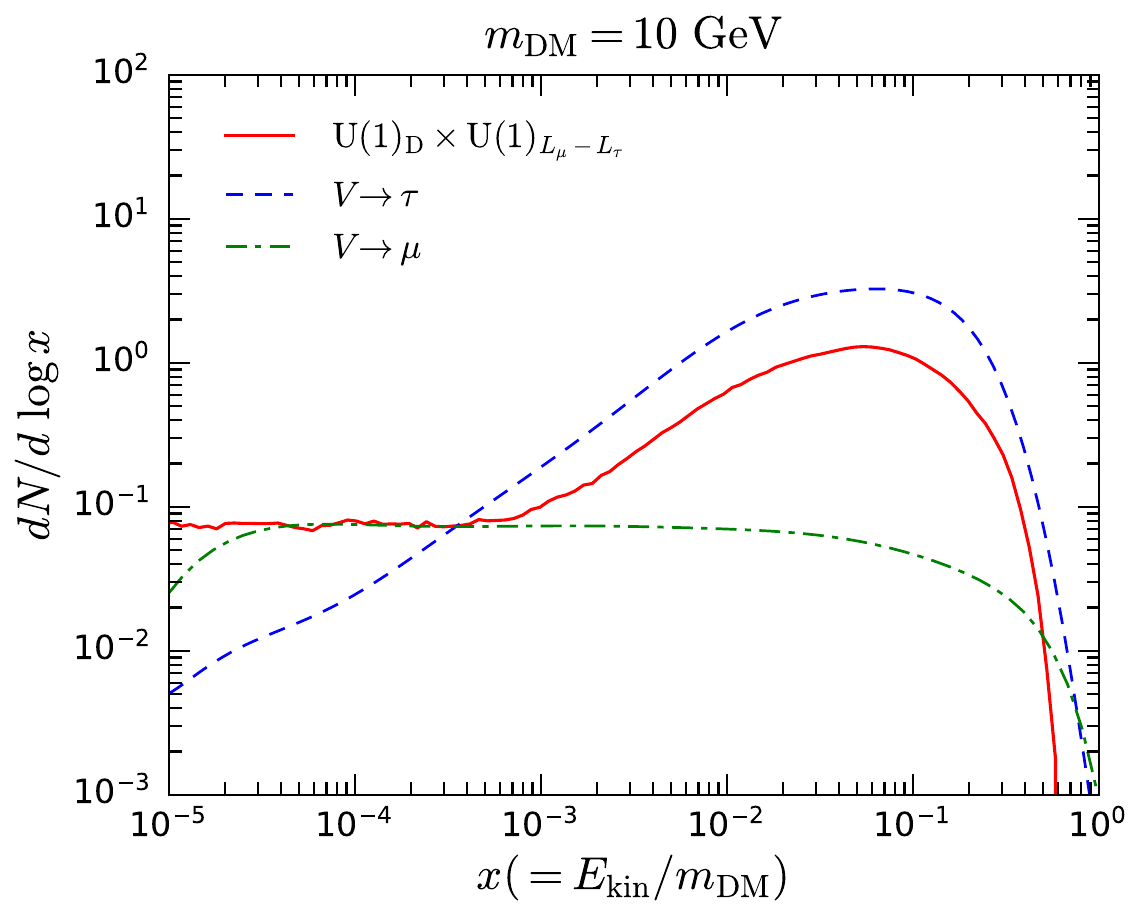}}
\caption{Gamma-ray spectra for $10~\mathrm{GeV}$ DM annihilating through various channels. Red solid lines represent the total spectra for the $\UoneD \times \UoneY$ model (left panel) and the $\UoneD \times \Uone_{L_\mu-L_\tau}$ model (right panel). Blue dashed lines show the spectra obtained from $b\bar{b}$ channel (left panel) and from the $V\to \tau$ channel (right panel), while green dotted-dashed lines represent the spectra from the $V\to \mu$ channel in both panels. }
\label{spec2}
\end{figure}

To better demonstrate how hadronic and leptonic final states shape the energy spectrum, we show a comparison of spectra between these two models and some single annihilation channels in Fig.~\ref{spec2}. The energy spectrum of the $\UoneD \times \UoneY$ model exhibits similar characteristics as the $b\bar{b}$ channel at high energies, reflecting its hadronic decay channels, while mimicking the $V \to \mu$ channel at low energies due to its leptonic decay channels~\footnote{$V \to \mu$ and $V \to \tau$ represent the processes $2\mathrm{DM}\to 2A' \to 4\mu$ and $2\mathrm{DM}\to 2A' \to 4\tau$, respectively}. In the $\UoneD \times \Uone_{L_\mu-L_\tau}$ model, since the $A'$ boson decays exclusively into $\mu$ pairs, $\tau$ pairs, and their corresponding neutrinos, the resulting energy spectrum manifests as a superposition of the $V \to \mu$ and $V \to \tau$ channels. Note that the spectrum from $V \to \tau$ in high energy region is similar to the $b\bar{b}$ channel, which is caused by the hadronic decay processes of tau lepton.

After calculating the gamma-ray flux for the above two models, we now proceed to perform a joint analysis of 42 dSphs using the likelihood functions provided by the Fermi-LAT collaboration. Following the approach outlined in~\cite{McDaniel:2023bju}, the likelihood function of J-factor is given by,
\begin{eqnarray}
\mathcal{L}(J_i|J_{\mathrm{obs},i}, \sigma_i) = \frac{1}{\ln(10)J_{\mathrm{obs},i}\sqrt{2\pi}\sigma_i}e^{-\frac{\left( \log_{10}(J_i)-\log_{10}(J_{\mathrm{obs},i}) \right)^2}{2\sigma_i^2}}, 
\end{eqnarray}
where $J_i$ represents the true J-factor value for the dSph $i$, while $J_{\mathrm{obs},i}$ and $\sigma_i$ denote the observed J-factor and its associated statistical uncertainty, respectively. The values for $J_{\mathrm{obs},i}$ and $\sigma_i$ are adopted from the analysis presented in~\cite{McDaniel:2023bju}, which follows the study in~\cite{Pace:2018tin, DES:2019vzn}. The J-factors of the dSphs and their uncertainties used in our analysis are listed in Table~\ref{tab:dSphs_J_factors} in Appendix~\ref{J-factors}. 

\begin{figure}[h!]
\centering
\subfigure[\label{Ind_dec_3-1}]
{\includegraphics[width=0.48\textwidth]{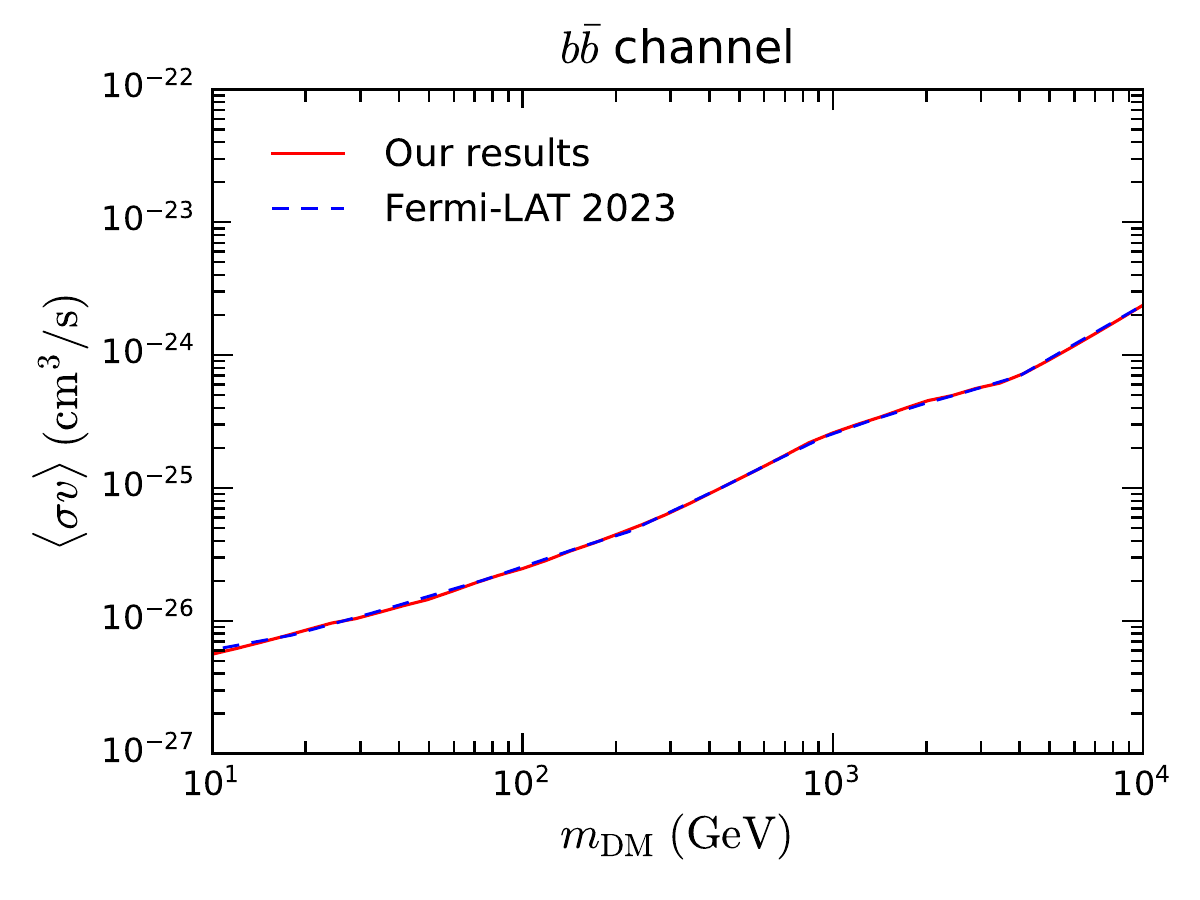}}
\hspace{.01\textwidth}
\subfigure[\label{Ind_dec_3-2}]
{\includegraphics[width=0.48\textwidth]{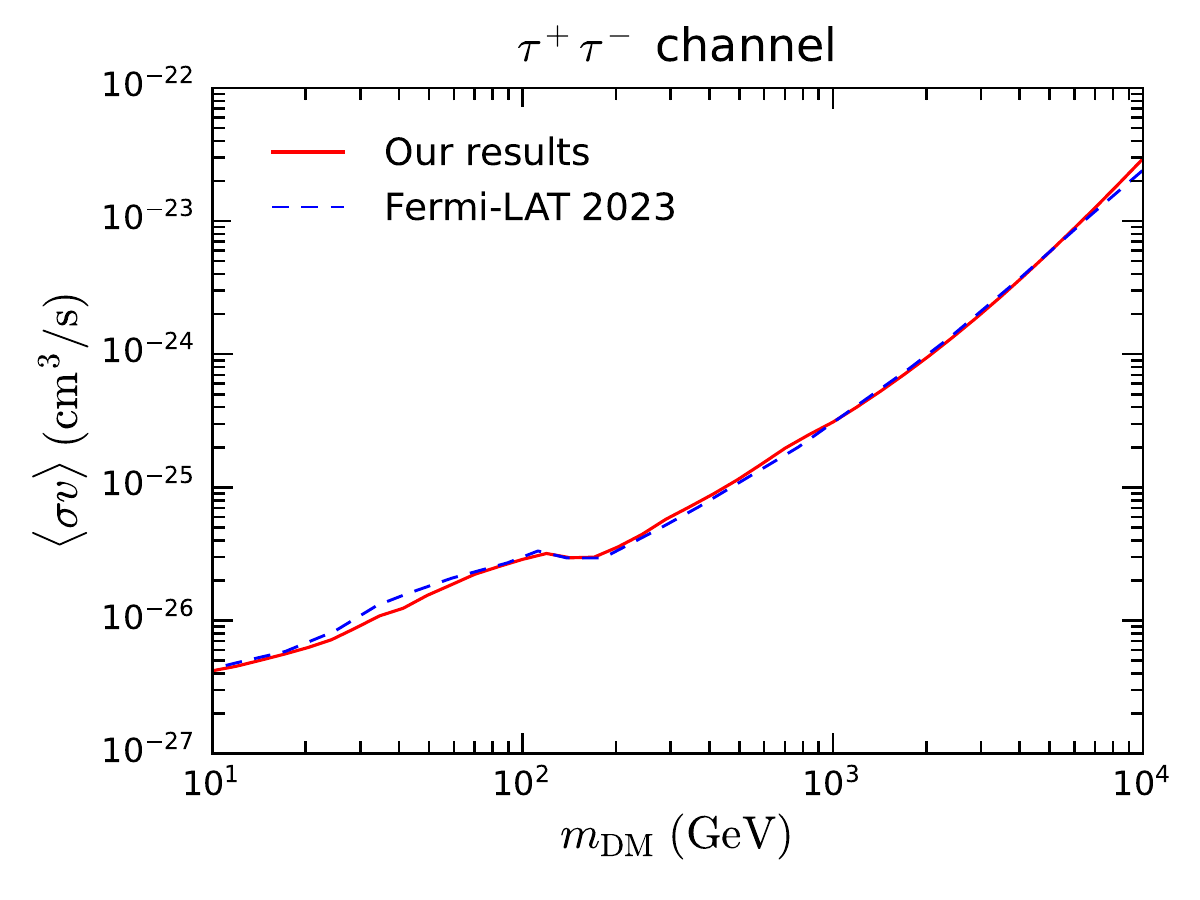}}
\caption{ A comparison of the $95\%$ CL upper limits on $\langle \sigma v \rangle$ for the $b\bar{b}$ (left panel) and $\tau^+ \tau^- $ (right panel) channels we obtained (red solid lines), with the results reported by the Fermi-LAT collaboration (blue dashed lines)~\cite{McDaniel:2023bju}.}
\label{Ind_dec_3}
\end{figure}

Therefore, we can construct the joint likelihood function as follows:
\begin{eqnarray}
\mathcal{L}(\langle \sigma v \rangle; \bm{v}|\bm{\mathcal{D}}) = \prod_i \mathcal{L}(\langle \sigma v \rangle; J_i, \bm{\mu}_i|\bm{\mathcal{D}}) \cdot \mathcal{L}(J_i|J_{\mathrm{obs},i}, \sigma_i),
\end{eqnarray}
where $\bm{v}$ represents the nuisance parameters, $\bm{\mu}_i$ includes any additional nuisance parameter other than $J_i$, and $\bm{\mathcal{D}}$ denotes the gamma-ray data. We then perform a test statistic (TS) analysis to determine the one-sided, $95\%$ confidence level (CL) upper limits on $\langle \sigma v \rangle$. This is achieved by identifying a decrease of $2.71/2$ in the log-likelihood from its maximum value. 
To validate our analysis method, we reproduce the $95\%$ CL limits for the $2\mathrm{DM}\to b\bar{b}$ and $2\mathrm{DM}\to\tau^+ \tau^-$ channels, as shown in Fig.~\ref{Ind_dec_3}. We find an excellent agreement between our results and the limits reported by the Fermi-LAT collaboration.

\subsection{Cosmic Microwave Background}
\label{CMB}

DM annihilation can inject electromagnetically interacting particles into the thermal bath in early universe, which might affect the anisotropies of Cosmic Microwave Background (CMB) observed by experiments like Planck~\cite{Planck:2015fie, Planck:2018vyg}. This effect is very significant during the cosmic dark ages (corresponding to redshifts $z \sim 20-1000$), the period after recombination but before reionization. Photons and $e^+ e^-$ pairs produced by DM annihilation, can ionize the ambient hydrogen gas and heat the plasma. This increases the residual ionization fraction, allows free electrons to scatter the CMB photons, and thus broadens the last scattering surface~\cite{Adams:1998nr, Chen:2003gz, Padmanabhan:2005es}. Therefore, the high precision CMB measurements place robust constraints on any energy injection from DM annihilation.

The energy deposited into the gas from DM annihilation can be characterized by a redshift- and model-dependent factor $f(z)$, which is defined as the deposited energy normalized to the energy injected at the same redshift. A portion of this deposited energy contributes to the ionization of the gas, and the final ionizing energy can be expressed as the product of $f(z)$ and the fraction of deposited energy going into ionization. Notably, uncertainties in calculating this fraction can be absorbed into $f(z)$, resulting in a corrected $f(z)$ curve. This is also the approach adopted by the Planck Collaboration. 

However, determining $f(z)$ is complicated as it requires tracking the energy loss and interaction processes of high-energy injected particles. A significant advance was made in~\cite{Slatyer:2015kla}, which provides corrected $f(z)$ curves for DM annihilation into photons and $e^+ e^-$ pairs. Performing a principal component analysis (PCA) on $f(z)$ curves, a universal weighting function $W(z)$ was derived to convert $f(z)$ into an effective deposition efficiency $f_{\mathrm{eff}}(E)$ for photons and $e^+ e^-$ pairs respectively, and thus the analysis of DM annihilation effects on the CMB is greatly simplified.

\begin{figure}[!h]
\centering
\subfigure[\label{CMB_1-1}]
{\includegraphics[width=0.48\textwidth]{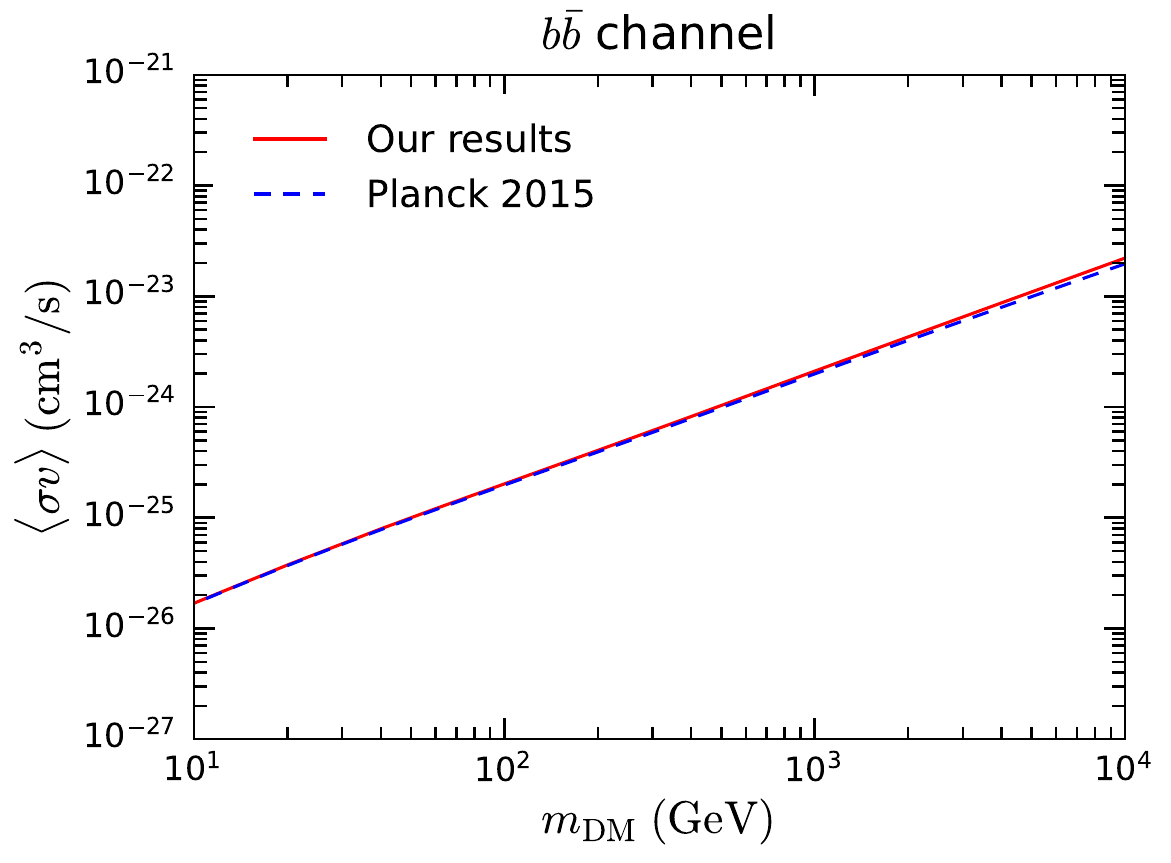}}
\hspace{.01\textwidth}
\subfigure[\label{CMB_1-2}]
{\includegraphics[width=0.48\textwidth]{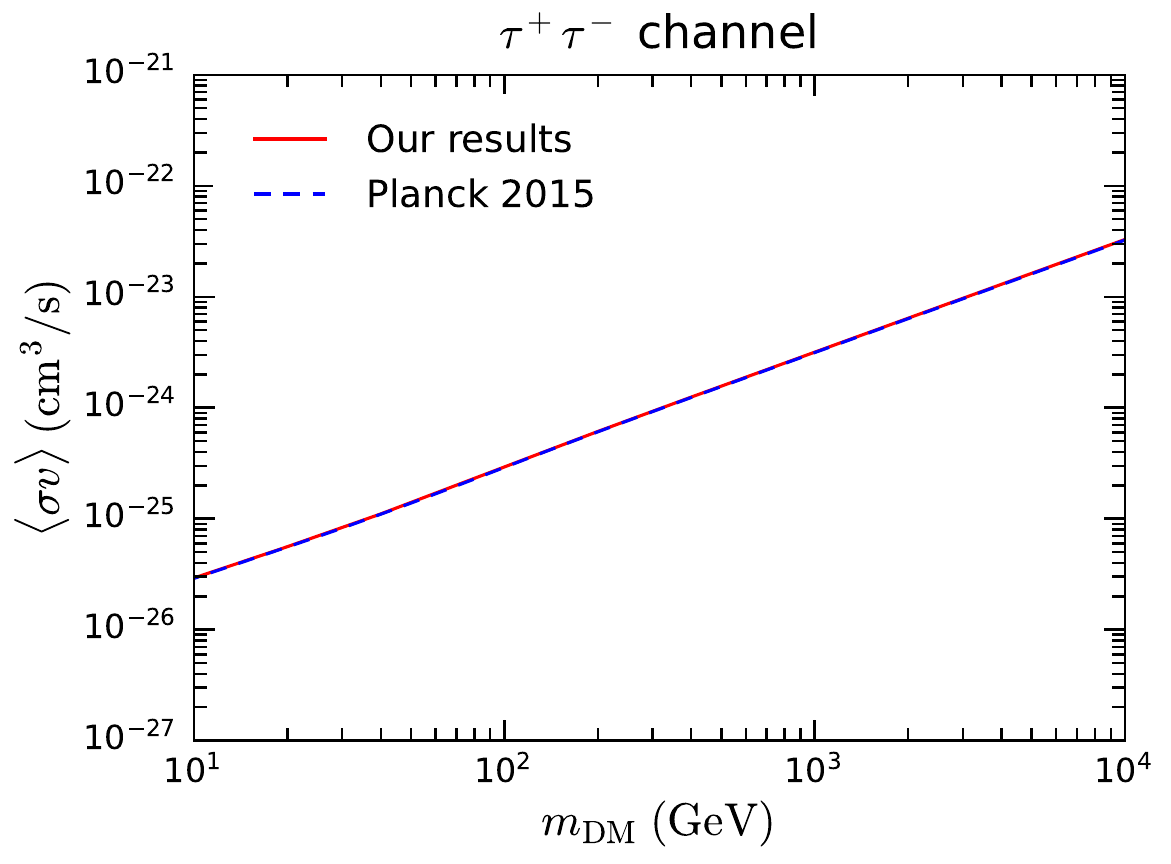}}
\caption{ Comparison of our derived $95\%$ CL upper limits (red) from Planck 2015 result on $\langle \sigma v \rangle$ as a function of $m_{\mathrm{DM}}$ for DM annihilation into $b\bar{b}$ (left panel) and $\tau^+ \tau^- $ (right panel) with those given by~\cite{Slatyer:2015jla} (blue).}
\label{CMB_1}
\end{figure}

Given the energy spectra of positrons $(dN/dE)_{e^+}$ and photons $(dN/dE)_{\gamma}$, the weighted $f_{\mathrm{eff}}(m_\mathrm{DM})$ can be calculated as~\cite{Slatyer:2015jla}:
\begin{eqnarray}
f_{\mathrm{eff}}(m_\mathrm{DM}) = \frac{1}{2 m_\mathrm{DM}} \int_0^{m_\mathrm{DM}} \, EdE \left[  2f_{\mathrm{eff}}^{e^+ e^-}(E)\left(\frac{dN}{dE}\right)_{e^+} + f_{\mathrm{eff}}^{\gamma}(E)\left(\frac{dN}{dE}\right)_{\gamma} \right].
\end{eqnarray}
The impact of DM annihilations on the CMB can then be characterized by:
\begin{eqnarray}
p_{\mathrm{ann}} = f_{\mathrm{eff}} \frac{\langle \sigma v \rangle}{m_\mathrm{DM}}.
\end{eqnarray}
In 2018, the Planck collaboration has established a $95\%$ C.L. upper limit on $p_{\mathrm{ann}}$ after marginalizing over other cosmological parameters~\cite{Planck:2018vyg}:
\begin{eqnarray}\label{pann}
p_{\mathrm{ann}} < 3.2 \times 10^{-28}~\mathrm{cm^3 s^{-1} GeV^{-1}}.
\end{eqnarray}
This bound shows considerable improvement compared to the Planck 2015 result, $p_{\mathrm{ann}} < 4.1 \times 10^{-28}~\mathrm{cm^3 s^{-1} GeV^{-1}}$, primarily due to an enhanced understanding and treatment of polarization systematics in the Planck polarization spectra. To crosscheck our methods with those presented in~\cite{Slatyer:2015jla}, we reproduce the CMB constraints on various DM annihilation channels. Fig.~\ref{CMB_1} presents these constraints for the $b\bar{b}$ (left panel) and $\tau^+ \tau^-$ (right panel) channels, showing excellent agreement as well. Note that Ref.~\cite{Slatyer:2015jla} used the constraints from Planck 2015 data, whereas our analysis in Sect.~\ref{results} employs the constraint (\ref{pann}) from Planck 2018 data.

\section{Results}
\label{results}

In this section, we present the DM relic density, Fermi-LAT and CMB constraints for the $\UoneD \times \UoneY$-portal and $\UoneD \times \Uone_{L_\mu-L_\tau}$-portal models. Moreover, we apply these constraints to the three types of DM introduced in Sect.~\ref{model}.

\begin{figure}[h!]
\centering
\subfigure[\label{Results_1-1}]
{\includegraphics[width=0.48\textwidth]{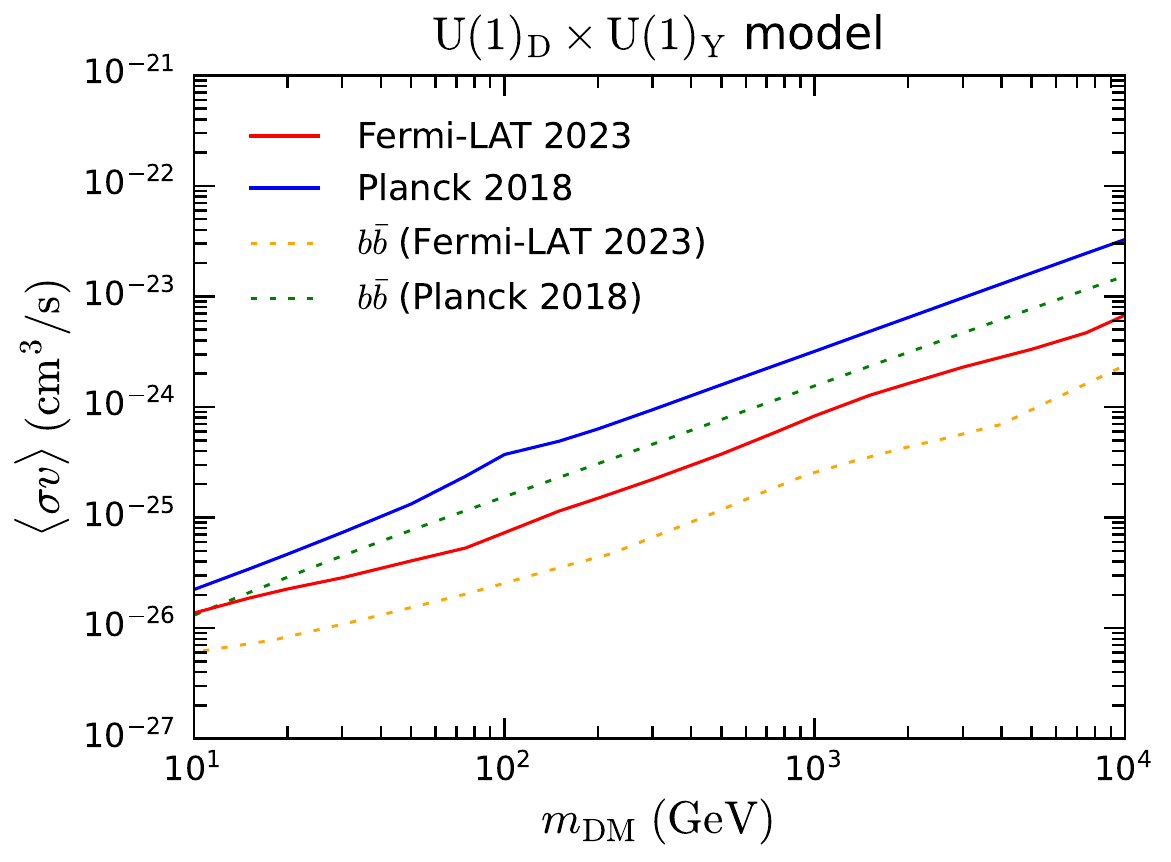}}
\hspace{.01\textwidth}
\subfigure[\label{Results_1-2}]
{\includegraphics[width=0.48\textwidth]{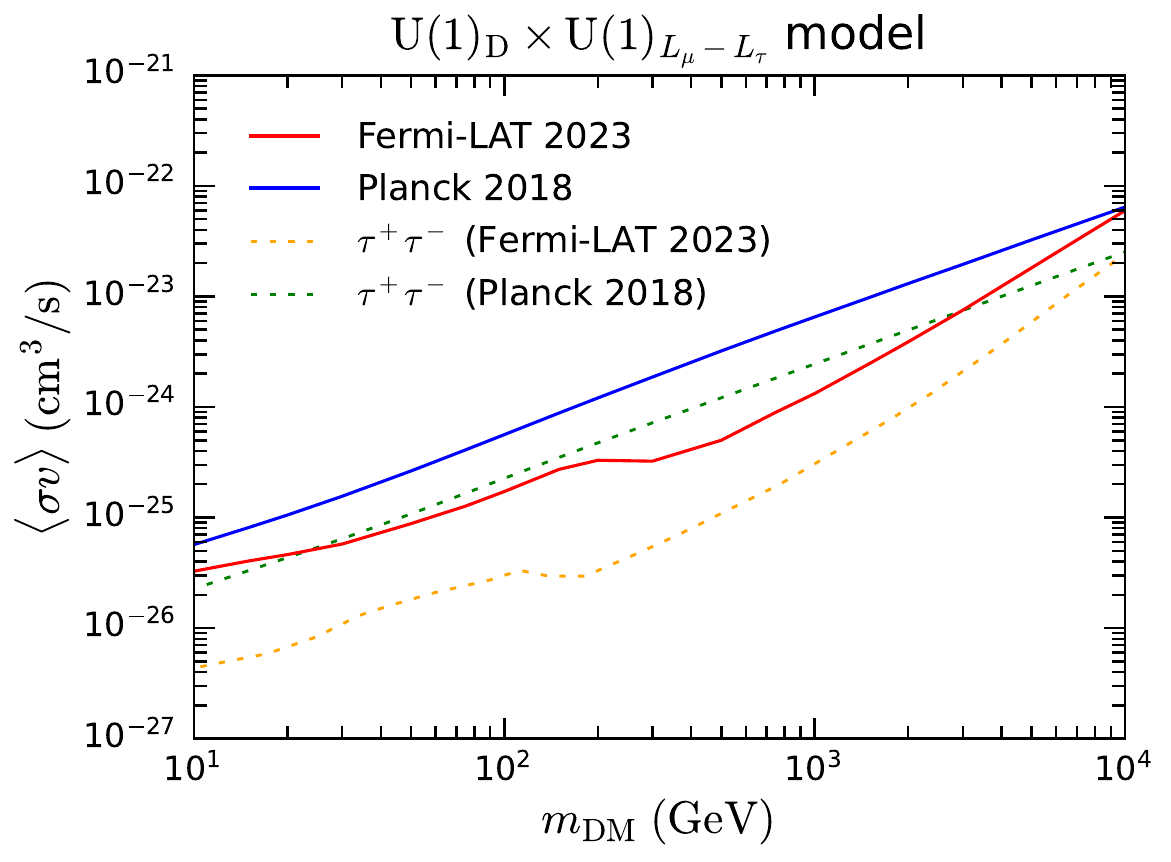}}
\caption{ $95\%$ CL upper limits on $\langle \sigma v \rangle$ as a function of $m_{\mathrm{DM}}$ from Fermi-LAT 2023 (red) and Planck 2018 (blue) for the $\UoneD \times \UoneY$ (left) and $\UoneD \times \Uone_{L_\mu-L_\tau}$ (right) models. Constraints for simplified $b\bar{b}$-channel (left) and $\tau^+ \tau^-$-channel (right) models from Fermi-LAT (orange) and Planck (green) are also included for comparison}.
\label{Results_1}
\end{figure}

In Fig.~\ref{Results_1}, we provide the constraints on $\langle \sigma v \rangle$ as a function of $m_{\mathrm{DM}}$, which are valid for all spins of DM. The Fermi-LAT 2023 constraints, obtained from a combined analysis of 42 dSphs, are indicated by red solid lines. The Planck 2018 constraints on energy injection into CMB from DM annihilation are indicated by blue solid lines. As a comparison, constraints derived from the assumption of single channel $2\mathrm{DM}\to b\bar{b}$ (left panel) or $2\mathrm{DM}\to\tau^+ \tau^-$ (right panel) are also presented. Orange dashed lines represent the constraints from Fermi-LAT, while the green dashed lines represent the constraints from the Planck experiment. The updated Planck constraints for these channels are slightly tighter than those presented in Fig.~\ref{CMB_1} adopted from the Planck 2015 result.

Both panels in Fig.~\ref{Results_1} show that the constraints from the Fermi-LAT (gamma rays) are more stringent than those from the Planck (CMB) across the entire mass range. As shown in Fig.~\ref{Results_1-1}, the Fermi-LAT bounds for the $\UoneD \times \UoneY$ model are looser than the simplified $b\bar{b}$-channel model in which the final states are purely hadronic. According to Fig.~\ref{Results_1-2}, we can see that the $\UoneD \times \Uone_{L_\mu-L_\tau}$ model performs much better than the simplified $\tau^+ \tau^-$-channel model since parts of the four-leptons final states consist of neutrinos, which do not radiate photons.

\begin{figure}[h!]
\centering
\subfigure[\label{Results_2-1}]
{\includegraphics[width=0.48\textwidth]{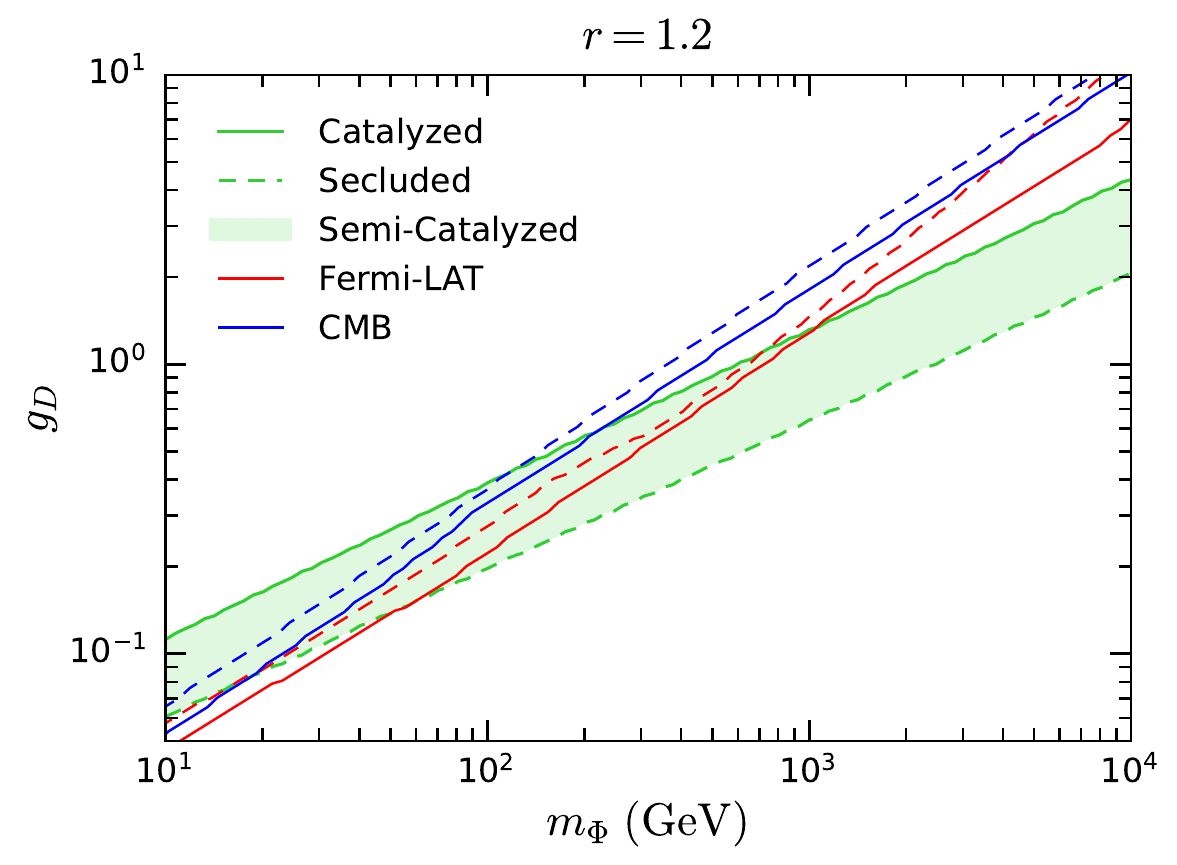}}
\hspace{.01\textwidth}
\subfigure[\label{Results_2-2}]
{\includegraphics[width=0.48\textwidth]{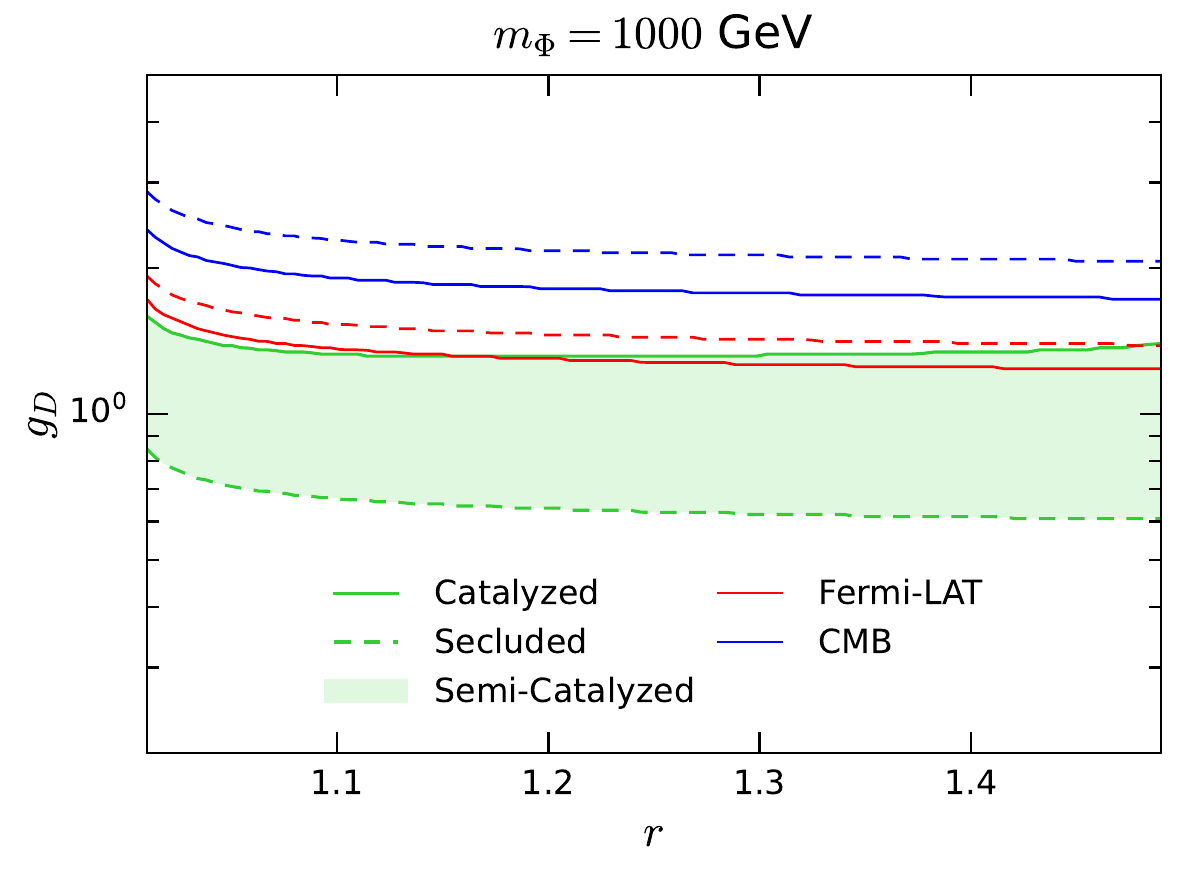}}
\hspace{.01\textwidth}
\subfigure[\label{Results_2-3}]
{\includegraphics[width=0.48\textwidth]{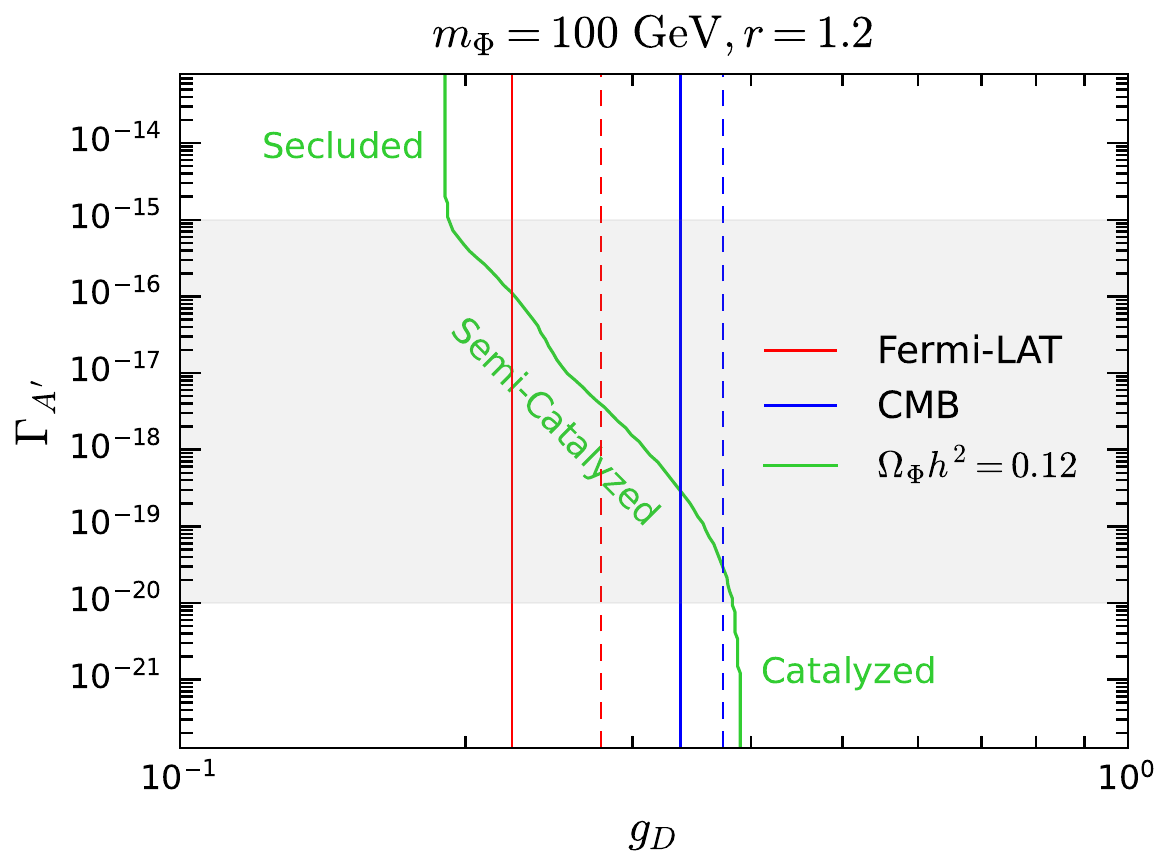}}
\caption{ Parameter spaces for complex scalar DM. The first panel shows $g_D$ versus $m_\Phi$ with $r = 1.2$, while the second panel shows $g_D$ versus $r$ with $m_{\Phi} = 1000~\mathrm{GeV}$. The last panel shows the relationship between $\Gamma_{A'}$ and $g_D$ for $m_\Phi = 100~\mathrm{GeV}$ and $r = 1.2$. The green areas (upper panels) and the green curve (below panel) represent the region satisfying $\Omega_{\Phi} h^2 = 0.12$. The red and blue curves represent the Fermi-LAT and CMB constraints, with the solid and dashed ones corresponding to the $\UoneD \times \UoneY$ and $\UoneD \times \Uone_{L_\mu-L_\tau}$ models, respectively.}
\label{Results_2}
\end{figure}

In Fig.~\ref{Results_2}, we illustrate the upper limits on the gauge coupling $g_D$ for the complex scalar DM model under three different setups of fixed parameters. Gamma-ray (red) and CMB (blue) bounds are depicted, with solid and dashed lines corresponding to the $\UoneD \times \UoneY$ and $\UoneD \times \Uone_{L_\mu-L_\tau}$ models, respectively. Fig.~\ref{Results_2-1} shows the parameter space of $g_D$ versus $m_\Phi$ with $r=1.2$. The green band denotes the parameter space in which the correct DM relic density $\Omega_{\Phi} h^2 = 0.12$ can be achieved. The green dashed boundary of this region represents the scenario of catalyzed annihilation DM, while the green solid boundary represents the scenario of secluded DM. The intermediate area represents the scenario of semi-catalyzed annihilation, in which the catalyzed annihilation processes are terminated by the decay of $A'$ leading to a sudden freeze-out of DM. The Fig.~\ref{fig5} shown in the previous section, can be regarded as a vertical slice at $m_{\Phi} = 1000~\mathrm{GeV}$, which shows a transition from the catalyzed case to the secluded case as $s_\epsilon$ increases, under the assumption of reproducing the correct relic density. Outside the green band, the correct relic density cannot be achieved in any case. Additionally, the gamma-ray and CMB constraints for $\UoneD \times \Uone_{L_\mu-L_\tau}$ model are milder than the $\UoneD \times \UoneY$ model, as we expect. Specifically, in the $\UoneD \times \Uone_{L_\mu-L_\tau}$ model, DM mass below $\sim 709~\mathrm{GeV}$ is excluded in the catalyzed annihilation scenario, whereas in the $\UoneD \times \UoneY$ model, DM mass has a lower bound $\sim 1052~\mathrm{GeV}$. For the secluded scenario, the constraints on the DM mass are relaxed to $\sim 16~\mathrm{GeV}$ and $\sim 59~\mathrm{GeV}$, respectively. Overall, the constraint for the secluded scenario is weaker than those on catalyzed scenario since the latter usually requires a stronger coupling $g_D$. The constraints for the semi-catalyzed scenario fall within them, depending on the value of $s_\epsilon$ ($\Gamma_{A'}$).

Fig.~\ref{Results_2-2} shows the parameter space of $g_D$ versus $r$, with $m_{\Phi}$ to be fixed at $1000~\mathrm{GeV}$. As $r$ increases, both the green band and the experiment limits slightly shift downward due to an increase in the annihilation cross section $\langle \sigma_2 v \rangle$, which requires a smaller $g_D$ to maintain the correct relic density. For the secluded scenario, the $g_D$ implied by the relic density is far below the Fermi-LAT and CMB bounds in both models. However, for the catalyzed annihilation scenario, the region of $r\gtrsim 1.16(1.48)$ is excluded by the Fermi-LAT data. Note that the curve implied by DM relic density exhibits a different dependence on $r$ compared to the limit curve derived from the Fermi-LAT data. This difference arises because the former is determined by both $\langle \sigma_2 v \rangle$ and $\langle \sigma_3 v^2 \rangle$, whereas the latter depends solely on $\langle \sigma_2 v \rangle$. This results in the two curves intersecting at an intermediate point and yields an upper limit of $r$ for the catalyzed annihilation scenario.

The last panel in Fig.~\ref{Results_2} shows the relationship between the decay width $\Gamma_{A'}$ and $g_D$ for fixing $m_\Phi = 100~\mathrm{GeV}$ and $r=1.2$. The green line represents the parameters which achieve $\Omega_{\Phi} h^2 = 0.12$. As $\Gamma_{A'}$ increases, DM transits from the catalyzed annihilation case to a secluded case. In the semi-catalyzed region, $\Gamma_{A'}$ and $g_D$ demonstrate an inverse relationship, consistent with Fig.~\ref{fig5}. In the secluded and catalyzed regions, further increases or decreases in $\Gamma_{A'}$ no longer influence the final relic density, making the green line vertical to the $g_D$ axis in these regions. Overall, for $100~\mathrm{GeV}$ DM, $g_D$ must lie between $0.19$ and $0.39$ to ensure the correct relic density. Constraints from gamma rays and CMB are also presented. For $\UoneD \times \UoneY$ and $\UoneD \times \Uone_{L_\mu-L_\tau}$ models with fixing $m_\Phi = 100~\mathrm{GeV}$ and $r=1.2$, the gauge coupling is restricted to be $g_D < 0.22$ ($0.28$) and $0.34$ ($0.37$) by Fermi-LAT and Planck data, respectively.

\begin{figure}[h!]
\centering
\subfigure[Fermion DM ($r=1.2$)\label{Results_3-1}]
{\includegraphics[width=0.48\textwidth]{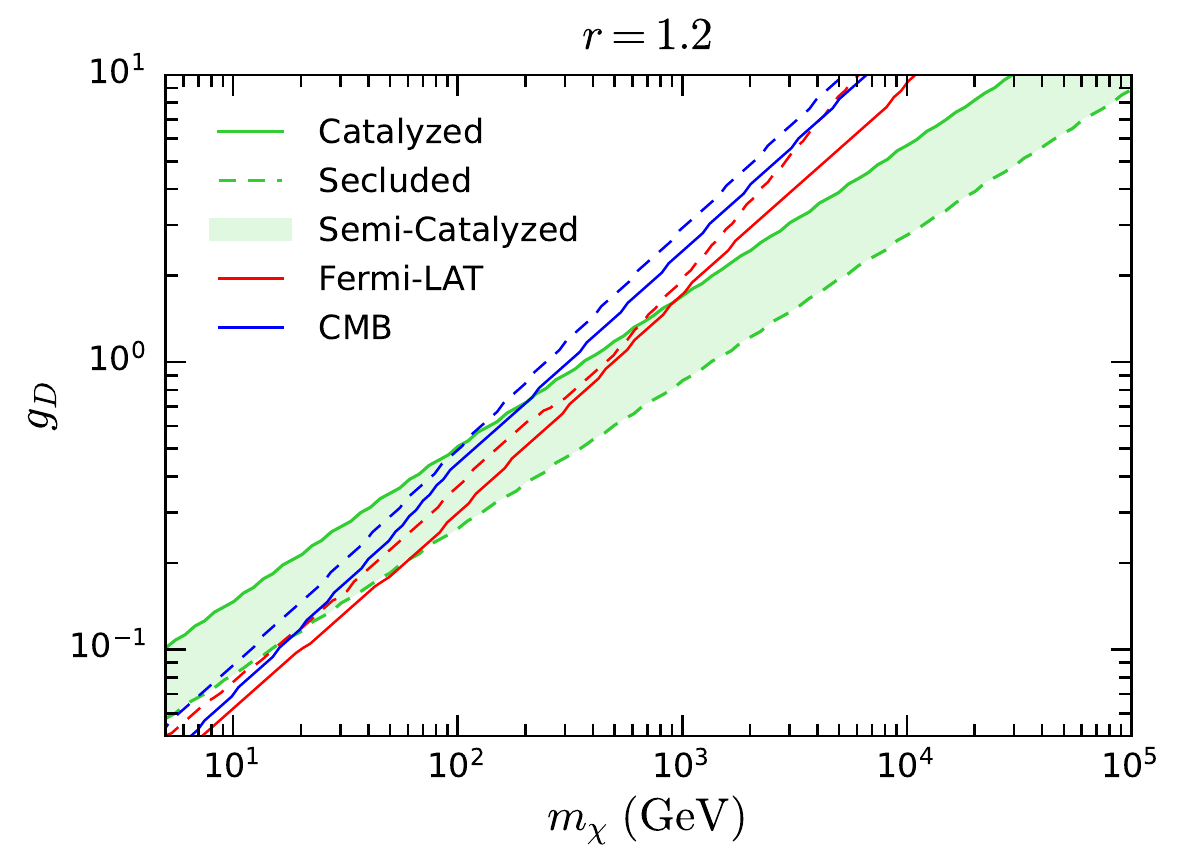}}
\hspace{.01\textwidth}
\subfigure[Fermion DM ($r=1.45$)\label{Results_3-2}]
{\includegraphics[width=0.48\textwidth]{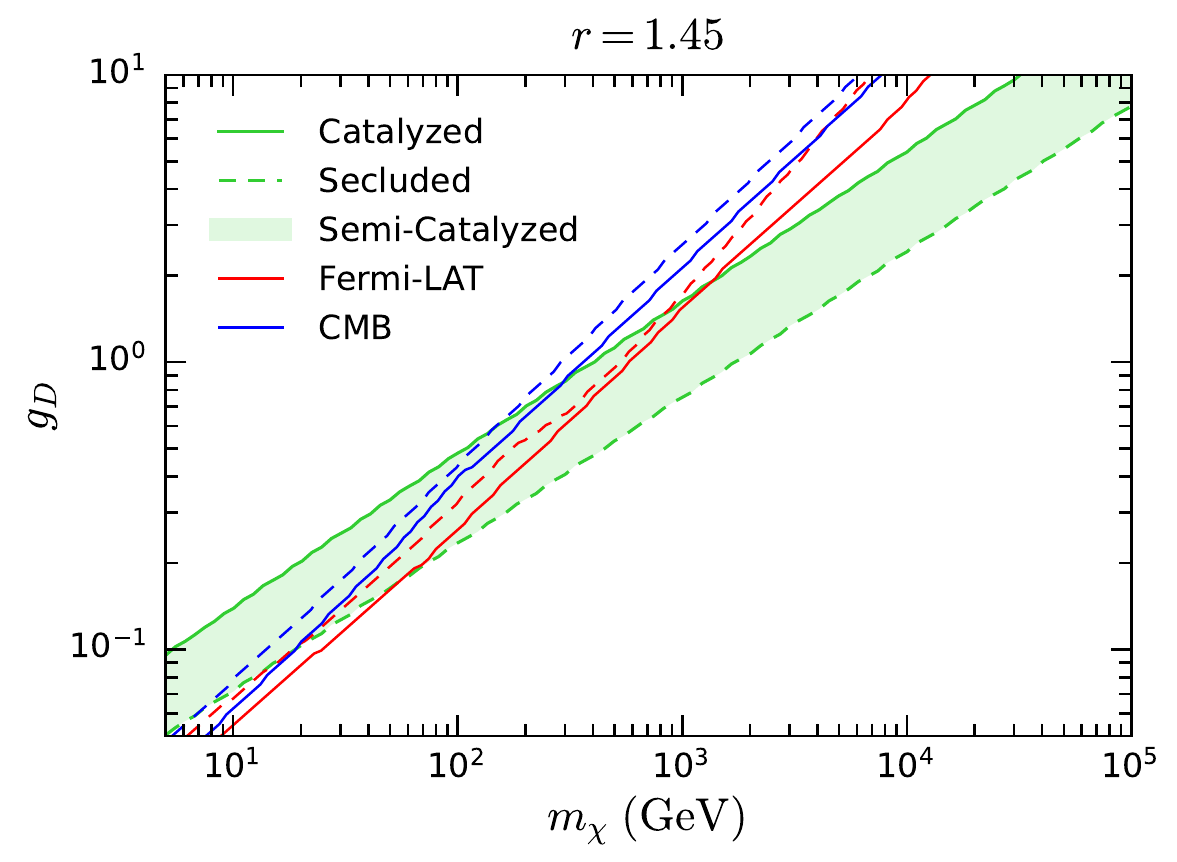}}
\hspace{.01\textwidth}
\subfigure[Vector DM\label{Results_3-3}]
{\includegraphics[width=0.48\textwidth]{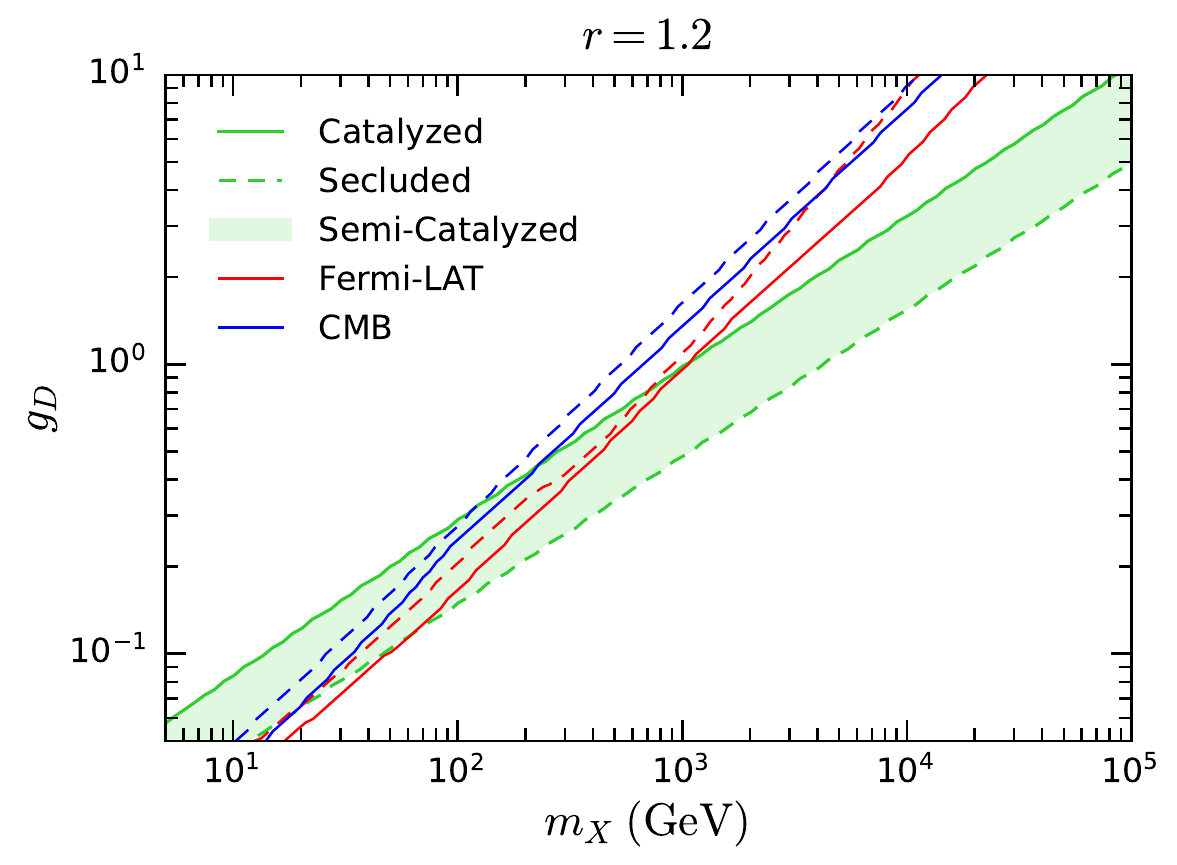}}
\caption{Fermi-LAT and CMB constraints for fermionic and vector DM.}
\label{Results_3}
\end{figure}

Finally, we present the gamma-ray and CMB constraints for fermionic DM (Fig.~\ref{Results_3-1}) and vector DM (Fig.~\ref{Results_3-3}) DM with fixing $r=1.2$ as a benchmark. For the Dirac fermion case, in the catalyzed annihilation scenario, the lower limits on the DM mass are $\sim 910~\mathrm{GeV}$ in the $\UoneD \times \UoneY$ and $\sim 665~\mathrm{GeV}$ in the $\UoneD \times \Uone_{L_\mu-L_\tau}$ model. In the secluded DM scenario, these limits are determined to be $\sim 61~\mathrm{GeV}$ and $\sim 17~\mathrm{GeV}$. In order to compare the results obtained through $2\mathrm{DM}\to 2A'\to 4\mathrm{SM}$ channels and those obtained in previous works using the simplified $2\mathrm{DM}\to 2\mathrm{SM}$ channel, we fix the parameter $r$ at $r = 1.45$ and show the limits in Fig.~\ref{Results_3-2}. In Ref.~\cite{Xing:2021pkb}, which considered the $\UoneD \times \UoneY$ model, the Fermi-LAT limits based on $2\mathrm{DM}\to 2\mathrm{SM}$ channel analysis are presented. The results adopted from their Fig.~4 showed the bounds for the DM mass to be $m_\chi \gtrsim 2.7~\mathrm{TeV}$ (catalyzed) and $m_\chi \gtrsim 100~\mathrm{GeV}$ (secluded) at $r \sim 1.5$. However, our analysis based on $2\mathrm{DM}\to 2A'\to 4\mathrm{SM}$ channels shows that the limits should be modified to $m_\chi \gtrsim 1.4~\mathrm{TeV}$ (catalyzed) and $m_\chi\gtrsim 55~\mathrm{GeV}$ (secluded), which are considerably relaxed. The limits for the $\UoneD \times \Uone_{L_\mu-L_\tau}$ model are even further relaxed, since there is less hadronic decay in this case. Similar conclusions hold for the CMB constraints as well. These results are consistent with our conclusion drawn from the Fig.~\ref{Results_1}.

For the vector case (Fig.~\ref{Results_3-3}), in the catalyzed annihilation scenario, the lower limits on the DM mass is $1086~\mathrm{GeV}$ in the $\UoneD \times \UoneY$ and $706~\mathrm{GeV}$ in the $\UoneD \times \Uone_{L_\mu-L_\tau}$ model. In contrast, Ref.~\cite{Cai:2021wmu}, which is based on the limits obtained by $2\mathrm{DM}\to 2\mathrm{SM}$ channel, excludes the catalyzed annihilation vector DM from $\UoneD \times \UoneY$ model with a mass below $\sim 4.4~\mathrm{TeV}$ when $r=1.2$ (see their Fig.~4). The refined analysis based on $2\mathrm{DM}\to 2A'\to 4\mathrm{SM}$ channels in this work significantly relax the constraints.

\section{Conclusions}
\label{conclusions}
In this work, we have proposed a complex scalar dark matter model based on a hidden $\UoneD$ gauge symmetry. The complex scalar $\Phi$ charged under $\UoneD$ serves as the DM candidate while the new gauge boson $A'$ plays the role of mediator bridging the dark and the SM sectors via a kinetic mixing. We focus on a situation that the mass ratio of DM and $A'$, $r$, is within a range $1\lesssim r\lesssim1.5$. In this case, the most relevant annihilation processes for DM density evolution are $2\Phi \to 2A'$ and $3A' \to 2\Phi$. Depending on the decay width of $A'$, the freeze-out scenarios of DM can be various. To be precise, when $\Gamma_{A'}$ is small enough, the catalyzed annihilation processes proceed until DM freeze-out. On the other hand, if $\Gamma_{A'}$ is large enough, $A'$ maintains thermal equilibrium and $3A' \to 2\Phi$ can be neglected, resulting in a usual freeze-out scenario of the secluded DM model. Between these two limits, the catalyzed annihilation processes can proceed until $A'$ decays, so we call it a semi-catalyzed scenario.

We have calculated the annihilation cross sections for the $2\Phi \to 2A'$ and $3A' \to 2\Phi$ processes, and determined the DM phenomenology including relic density, indirect detection (gamma-ray detection of Fermi-LAT), and CMB (Planck). Since the dark sector couples to the SM sector feebly, the stringent direct detection bound can be easily circumvented, and thus the indirect searching is the most important way to detect DM. For comprehensiveness, we also consider the relic density and indirect detection constraints on the fermionic and vector DM models.

We have studied two models with different portals connecting the dark sector to SM: $\UoneD \times \UoneY$ and $\UoneD \times \Uone_{L_\mu-L_\tau}$. The former kinetically mixes $A'$ with the hypercharge gauge field $B$, while the latter mixes it with a new gauge boson $Z'$ originating from a $\Uone_{L_\mu-L_\tau}$ gauge symmetry. The $\UoneD \times \UoneY$ model enables $A'$ to decay into quarks and leptons, and thus more gamma rays are produced comparing to the $\UoneD \times \Uone_{L_\mu-L_\tau}$ model, which has less hadronic decay processes. We obtain the Fermi-LAT constraints from a combined analysis of 14.3 years of observations on 42 dSphs, and the CMB constraints from Planck experimental results of DM annihilation into photons and electron-positron pairs during the cosmic dark ages.

Finally, we present the relic density and indirect detection constraints for the $\UoneD \times \UoneY$ and $\UoneD \times \Uone_{L_\mu-L_\tau}$ models with different DM spins, and find viable parameter space for all cases. Since the dominant annihilation processes for DM are $2\mathrm{DM}\to 2A'\to 4\mathrm{SM}$ in our models, we find that both models have much weaker bounds on the $\langle\sigma v\rangle$ compared to the previous analysis based on the $2\mathrm{DM}\to 2\mathrm{SM}$ channels. Additionally, limits for $\UoneD \times \UoneY$ are more stringent than those for $\UoneD \times \Uone_{L_\mu-L_\tau}$ as we expected. For complex scalar DM with a mass ratio of $r=1.2$ between the DM and the mediator, a DM heavier than $709~\mathrm{GeV}$ ($16~\mathrm{GeV}$) in the catalyzed annihilation (secluded) scenario remains available under all constraints. For the fermionic and vector DM, our limits are more relaxed than previous studies using single-channel bounds. For instance, fermionic DM with $m_\chi\sim1~\mathrm{TeV}$ in the catalyzed annihilation scenario, which was excluded in the previous study, now remains available under the $\UoneD \times \Uone_{L_\mu-L_\tau}$ framework. For the vector DM with $r = 1.2$, the Fermi-LAT limit is relaxed from $\sim 4.4~\mathrm{TeV}$ to $706~\mathrm{GeV}$ in the catalyzed annihilation scenario.

\begin{acknowledgments}
C. C. is supported by the National Natural Science Foundation of China (NSFC) under Grant No. 11905300, and the Guangzhou Science and Technology Planning Project under Grant No. 2023A04J0008. H.-H. Z. is supported by the NSFC under Grant No. 12275367. This work is also supported by the
Fundamental Research Funds for the Central Universities, and the Sun Yat-Sen University Science Foundation.

\end{acknowledgments}

\appendix

\section{J-factors for the Benchmark Sample}
\label{J-factors}

\begin{longtable}{|l|c|c|c|c|}
\hline
\textbf{Name} & \textbf{R.A. (J2000)} & \textbf{Decl. (J2000)} & \textbf{Distance} & \textbf{log$_{10} J \pm \sigma_J$} \\
              & [deg]                 & [deg]                  & [kpc]             & [log$_{10}$GeV$^2$cm$^{-5}$] \\
\hline
Aquarius II   & 338.48  & -9.33   & 108.0  & 17.80 $\pm$ 0.55  \\
Boötes II     & 209.51  & 12.86   & 42.0   & 18.30 $\pm$ 0.95  \\
Canes Venatici I & 202.01 & 33.55  & 218.0  & 17.42 $\pm$ 0.16  \\
Canes Venatici II & 194.29 & 34.32 & 160.0  & 17.82 $\pm$ 0.47  \\
Carina        & 100.41  & -50.96  & 105.0  & 17.83 $\pm$ 0.10  \\
Carina II     & 114.11  & -58.0   & 36.0   & 18.25 $\pm$ 0.55  \\
Coma Berenices & 186.75 & 23.91   & 44.0   & 19.00 $\pm$ 0.35  \\
Draco         & 260.07  & 57.92   & 76.0   & 18.83 $\pm$ 0.12  \\
Draco II      & 238.17  & 64.58   & 22.0   & 18.93 $\pm$ 1.54  \\
Eridanus II   & 56.09   & -43.53  & 380.0  & 16.60 $\pm$ 0.90  \\
Fornax        & 39.96   & -34.5   & 147.0  & 18.09 $\pm$ 0.10  \\
Grus I        & 344.18  & -50.18  & 120.0  & 16.50 $\pm$ 0.80  \\
Hercules      & 247.77  & 12.79   & 132.0  & 17.37 $\pm$ 0.53  \\
Horologium I  & 43.88   & -54.12  & 79.0   & 19.00 $\pm$ 0.81  \\
Hydrus I      & 37.39   & -79.31  & 28.0   & 18.33 $\pm$ 0.36  \\
Leo I         & 152.11  & 12.31   & 254.0  & 17.64 $\pm$ 0.13  \\
Leo II        & 168.36  & 22.15   & 233.0  & 17.76 $\pm$ 0.20  \\
Leo IV        & 173.24  & -0.55   & 154.0  & 16.40 $\pm$ 1.08  \\
Leo V         & 172.79  & 2.22    & 178.0  & 17.65 $\pm$ 0.97  \\
Pegasus III   & 336.1   & 5.41    & 215.0  & 18.30 $\pm$ 0.93  \\
Pisces II     & 344.63  & 5.95    & 182.0  & 17.30 $\pm$ 1.04  \\
Reticulum II  & 53.92   & -54.05  & 30.0   & 18.90 $\pm$ 0.38  \\
Sagittarius II & 298.16 & -22.07  & 69.0   & 17.35 $\pm$ 1.36  \\
Segue 1       & 151.75  & 16.08   & 23.0   & 19.12 $\pm$ 0.53  \\
Sextans       & 153.26  & -1.61   & 86.0   & 17.73 $\pm$ 0.12  \\
Tucana II     & 342.98  & -58.57  & 58.0   & 18.97 $\pm$ 0.54  \\
Tucana IV     & 0.73    & -60.85  & 48.0   & 18.40 $\pm$ 0.55  \\
Ursa Major I  & 158.77  & 51.95   & 97.0   & 18.26 $\pm$ 0.28  \\
Ursa Major II & 132.87  & 63.13   & 32.0   & 19.44 $\pm$ 0.40  \\
Ursa Minor    & 227.24  & 67.22   & 76.0   & 18.75 $\pm$ 0.12  \\
Boötes IV     & 233.69  & 43.73   & 209.0  & 17.25 $\pm$ 0.60  \\
Carina III    & 114.63  & -57.79  & 28.0   & 19.70 $\pm$ 0.60  \\
Centaurus I   & 189.59  & -40.9   & 116.0  & 18.14 $\pm$ 0.60  \\
Cetus II      & 19.47   & -17.42  & 30.0   & 19.10 $\pm$ 0.60  \\
Cetus III     & 31.33   & -4.27   & 251.0  & 17.30 $\pm$ 0.60  \\
Columba I     & 82.86   & -28.01  & 183.0  & 17.60 $\pm$ 0.60  \\
Grus II       & 331.02  & -46.44  & 53.0   & 18.40 $\pm$ 0.60  \\
Phoenix II    & 355.0   & -54.41  & 83.0   & 18.30 $\pm$ 0.60  \\
Pictor I      & 70.95   & -50.29  & 114.0  & 18.00 $\pm$ 0.60  \\
Pictor II     & 101.18  & -59.9   & 46.0   & 18.83 $\pm$ 0.60  \\
Reticulum III & 56.36   & -60.45  & 92.0   & 18.20 $\pm$ 0.60  \\
Tucana V      & 354.35  & -63.27  & 55.0   & 18.90 $\pm$ 0.60  \\
\hline
\caption{Properties of 42 dSphs used in our analysis, including coordinates (R.A. and Decl.), distances, and J-factor values for each galaxy~\cite{McDaniel:2023bju}.} \label{tab:dSphs_J_factors}
\end{longtable}

\section{Kinetic Equilibrium Before DM Freeze-out}
\label{KE}

A template approach to maintaining the kinetic equilibrium of DM is to refer to the method proposed in section II.B of Ref.~\cite{Cline:2017tka}. We assume DM couples to a new scalar field $\Psi$, which is relativistic before the freeze-out of DM. $\Psi$ particle should be unstable and decay into SM particles after the freeze-out of DM. If we want DM to maintain kinetic equilibrium while ensuring that the evolution of its density is negligibly affected by this interaction, then the following conditions should be imposed:
\begin{enumerate}
\item\label{cond1} The interaction must be strong enough to maintain DM in kinetic equilibrium until freeze-out.
\item\label{cond2} The new annihilation cross section of $2\Phi \to 2\Psi$ must be sufficiently small to avoid altering the thermal evolution of DM described in Sect.~\ref{model}.
\end{enumerate}

\begin{figure}[!h]
\centering
\includegraphics[width=0.65\textwidth]{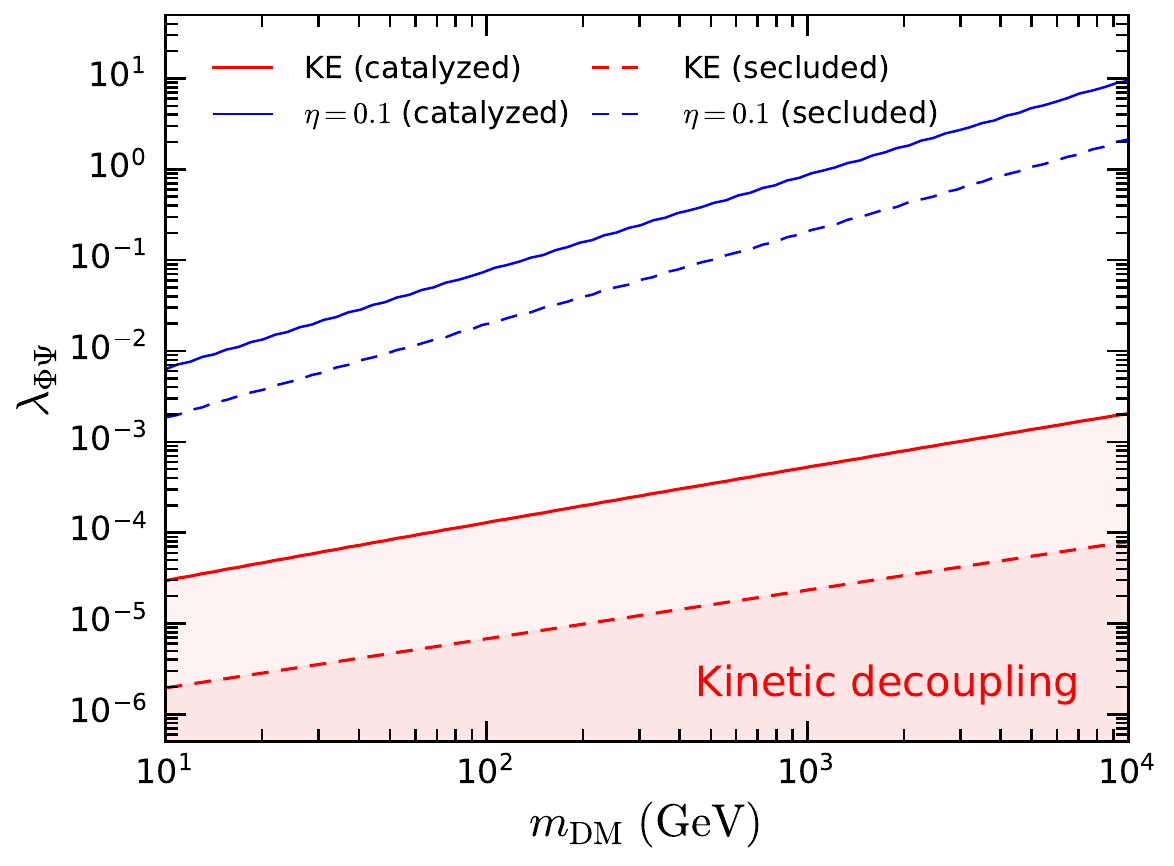}
\caption{Parameter space for $\lambda_{\Phi\Psi}$ versus $m_{\text{DM}}$, where the viable region lies between the KE condition bound (red curves) and $\eta = 0.1$ bound (blue curves) for both catalyzed annihilation (solid) and secluded (dashed) scenarios. }
\label{fig13}
\end{figure}

The interaction between $\Phi$ and $\Psi$ fields can be the following quartic term:
\begin{eqnarray}
\mathcal{L}_{\Phi \Psi} = \lambda_{\Phi \Psi} |\Phi|^2 |\Psi|^2.
\end{eqnarray}
The thermally averaged cross section for the processes $2\Phi \to 2\Psi$ and $\Phi\Psi \to \Phi\Psi$ can be easily derived as:
\begin{eqnarray}
\langle\sigma v\rangle_{2\Phi \to 2\Psi} = \frac{\lambda_{\Phi\Psi}^2}{32\pi m_\Phi^2},\quad\langle\sigma v\rangle_{\Phi\Psi \to \Phi\Psi} = \frac{\lambda_{\Phi\Psi}^2}{16\pi m_\Phi^2}
\end{eqnarray}
Then the \ref{cond1}st. condition can be estimated by~\cite{Hofmann:2001bi, Visinelli:2015eka, Cai:2021wmu},
\begin{eqnarray}\label{KE_condition}
\frac{T}{m_\Phi}\frac{n_\Psi \langle\sigma_{\Phi\Psi \to \Phi\Psi}\rangle}{H} \gtrsim 1,
\end{eqnarray}
where $H$ is the Hubble constant and $n_\Psi$ denotes the number density of $\Psi$. Fig.~\ref{fig13} illustrates the allowed parameter space for $\lambda_{\Phi\Psi}$ as a function of $m_{\text{DM}}$, satisfying both conditions. We fix the mass ratio between DM and mediator $A'$ at $1.2$, while $g_D$ varies with $m_{\text{DM}}$ to ensure the correct relic density. The regions above the red curves satisfy relation~\eqref{KE_condition}, with solid and dashed curves corresponding to the catalyzed annihilation and secluded scenarios, respectively. We define $\eta =\langle\sigma_{2}v\rangle_{2\Phi \to 2\Psi}/ \langle\sigma_{2}v\rangle_{2\Phi \to 2A'}$, with blue curves indicating $\eta = 0.1$. In the regions below the blue curves, the annihilation cross section for the $2\Phi \to 2\Psi$ process accounts for less than 10\% of the total quantity, so it is negligible. We find that it is easy to ensure the kinetic equilibrium of DM if $\lambda_{\Phi\Psi}\sim10^{-3}$ in both scenarios. For the fermionic and vector dark matter, the interactions between DM and $\Psi$ can be achieved by non-renormalizable operators (see Ref.~\cite{Cline:2017tka} as an example for the fermionic case). 

\bibliographystyle{utphys}
\bibliography{ref}

\end{document}